\documentclass[preprint,11pt,3p]{elsarticle}

\usepackage{graphicx}
\usepackage{epstopdf,epsfig}
\usepackage{amssymb}
\usepackage{amsmath}
\usepackage{makecell}
\usepackage{tikz}
\usepackage{dirtytalk}
\usepackage{float}
\usepackage{tabularx}
\usepackage{adjustbox}
\usetikzlibrary{shapes}
\usepackage{multirow}
\usepackage{xcolor,colortbl}
\usepackage{comment}

\DeclareRobustCommand\sampleline[1]{%
\tikz\draw[#1] (0,0) (0,\the\dimexpr\fontdimen22\textfont2\relax)
-- (1.6em,\the\dimexpr\fontdimen22\textfont2\relax);}

\usepackage{lineno}

\begin{document}

\begin{frontmatter}
\title{A disturbance corrected point-particle approach for two-way coupled particle-laden flows on arbitrary shaped grids}
\author[label1]{Pedram Pakseresht}
\address[label1]{School of Mechanical, Industrial and Manufacturing Engineering \\ Oregon State University \\ Corvallis, OR 97331, USA}
\ead{pakserep@oregonstate.edu}
\author[label1]{Sourabh V. Apte\corref{cor1}}
\ead{sourabh.apte@oregonstate.edu}
\cortext[cor1]{Corresponding author. 308 Rogers Hall, Corvallis, OR 97331, USA. Tel: +1 541 737 7335, Fax: +1 541 737 2600.}

\begin{abstract}
A general, two-way coupled, point-particle formulation that accounts for the disturbance created by the dispersed particles in obtaining the undisturbed fluid flow field needed for accurate computation of the force closure models is presented. Specifically, equations for the disturbance field created by the presence of particles are first derived based on the inter-phase momentum coupling force in a finite-volume formulation. Solution to the disturbance field is obtained using two approaches: (i) direct computation of the disturbance velocity and pressure using the reaction force due to particles at computational control volumes, and (ii) a linearized, approximate computation of the disturbance velocity field, specifically applicable for low Reynolds number flows. In both approaches, the computed disturbance field is used to obtain the undisturbed fluid velocity necessary to model the aerodynamic forces on the particle. The two approaches are thoroughly evaluated for a single particle in an unbounded and wall-bounded flow on uniform, anisotropic, as well as unstructured grids to show accurate computation of the particle motion and inter-phase coupling. The approach is straightforward and can be applied to any numerical formulation for particle-laden flows including Euler-Lagrange as well as Euler-Euler formulations.
\end{abstract}

\begin{keyword}
Undisturbed flow, Euler-Lagrange, Point-particle approach, Arbitrary grids.
\end{keyword}

\end{frontmatter}







\section{Introduction}
Particle-laden flows, wherein small size solid particles, liquid droplets or gaseous bubbles are dispersed in a fluid flow, are widely encountered in many engineering, biological and environmental applications. A wide range of numerical approaches resolving different scales of fluid and particle motion have been developed for accurate and predictive simulation of these flows~\citep{van2008numerical,balachandar2010_annualReview}. The point-particle (PP) approach~\citep{maxey1983},
in which particles are assumed spherical, subgrid and {\it modeled} as point sources, has received much attention in modeling particle-laden flows owing to its simplicity and affordability in simulating motion of large number of particles (${\mathcal O}(10^6-10^9)$). 

This approach was originally developed for dilute particulate flows with particles smaller than the fluid length scale (or the computational grid size) wherein the presence of particles does not significantly perturb the characteristics of the flow, i.e., one-way coupled regime~\citep{elghobashi1991}. In modeling this regime, the fluid phase equations are solved irrespective of the presence of the particles, while closures for fluid forces acting on particles such as drag, lift, added mass, pressure gradient, and history effect are employed to obtain the particle trajectories.  
Despite the original development of the PP approach for one-way coupled regimes, it has been widely applied to modeling two-way coupled regimes wherein the fluid phase is indeed perturbed by the presence of particles \citep{elghobashi1991}. Such a scenario may happen when the particle loading (or concentration) locally or globally becomes large, either due to a few large size particles or dense regime of small particles. The effect of particles on the carrier phase is then modeled by applying the particle reaction force to the fluid phase equations through a momentum source term. In addition, for dense loading, the volume occupied by the particles is also removed when solving for the fluid phase equations, by applying volume filtered equations, or volumetric coupling, which results in additional source terms in the continuity equation due to spatio-temporal variations in particle volume fraction. Unlike the standard approaches, this formulation couples two phases through both momentum and continuity equations~\citep{ferrante2004,apte2008,capecelatro2013,pakseresht2019}.

Since particles in PP approach are assumed to be smaller than the grid and the local flow over the particles is considered uniform, hence force closures based on uniform flow over a sphere are typically employed. However, to extend the applicability range of PP approach to particles of size on the order of or slightly larger than the computational grid, that receive spatially varying flow field, Faxen corrections have been developed~\citep{annamalai2017faxen}. 
Typical force closures that are derived for a single particle rely on the {\it undisturbed} fluid flow, which is not readily available in the two-way coupled simulations. By definition, the undisturbed flow is the velocity and pressure fields that would exist at the location of a particle if that particle was not there in the flow. For multi-particle systems, the undisturbed flow field seen by a particle only corresponds to the flow field in the absence of that specific particle, however, includes the disturbed flow created by the neighboring particles. In two-way coupled simulations, the fluid phase is altered by the self-induced disturbance of each individual particle through inter-phase momentum and mass exchange, and using the available disturbed flow field for force closures results in erroneous predictions.
The error introduced by using the disturbed field may remain small when the particle size is much smaller than the grid, owing to the negligible disturbance of small particles. However, when the particle size is on the order of or bigger than the grid resolution, such as those encountered in direct and large-eddy simulations, the disturbance due to particles becomes noticeable, hence the errors can become significantly large.

\cite{pan1996} first showed that the PP approach produces wrong prediction for velocity of a single particle settling toward a no-slip wall, and in order to improve the predictions, they introduced a velocity-disturbance model, wherein the analytical Stokes solution at the location of the particle is superimposed on the background flow to reflect the effect of the particle. Unlike PP approach, their model eliminates any dependency of the particle force computations to the undisturbed fluid velocity and results in more accurate predictions. Although their model can be applied to both unbounded and wall-bounded regimes due to the availability of the Stokes solution for both, it is limited to small particle Reynolds numbers ($Re_p$) and at steady state condition for which the analytical solution is available. 
\cite{gualtieri2015} regularized the PP approach for the unbounded flows by deriving analytical equations to remove the self-induced velocity disturbance created by the particles, that is also extended to wall-bounded regimes~\citep{battista2019}. Their approach requires considerable computational resources to resolve the stencil over which the particle force is distributed using a Gaussian filter function. \cite{horwitz2016,horwitz2018} originated a method to obtain the undisturbed field based on the enhanced curvature in the disturbed velocity field for particle Reynolds numbers of $Re_p{<}10.0$. A C-field library data was built using reverse engineering technique that needs to be added to the current PP packages for recovering the undisturbed velocity. 
Their model is limited to (i) isotropic rectilinear computational grids, (ii) flows with particles of maximum size of the grid, and (iii) unbounded regimes. 
\cite{ireland2017} derived an analytical expression for recovering the undisturbed velocity in the unbounded regimes based on the steady state Stokes solution that was derived as the solution of a feedback force distributed to the background flow using a Gaussian filter. Their model accounts for the displaced fluid mass by the particles and is limited to unbounded regimes with small $Re_p$.

Using analytical and empirical expressions, \cite{balachandar2019} developed a model that corrects the PP approach for cold particle-laden flows with $Re_p{<}200$, with its extension to heated particle-laden flows~\citep{liu2019self}, as well. Despite its applicability for a wide range of flow parameters, it is restricted to unbounded flows and a Gaussian filter function for projecting the particle's reaction force. Recently,~\citet{evrard2020euler} used Stokes flow through a regularised momentum source with extension to finite Reynolds numbers based on the Oseen flow solution using Green's functions, to obtain the undisturbed fluid velocity and showed good predictions for arbitrary particle-to-grid size ratio and a wide range of particle Reynolds numbers in an unbounded flow. \cite{esmaily2018} developed a generic correction scheme in which each computational cell, that is subjected to the particle force, is treated as a solid object that is immersed in the fluid and dragged at a velocity identical to the disturbance created by the particle. Although their physics-based model was devised to handle (i) relatively large size particles even larger than the grid resolution, (ii) isotropic and anisotropic grids, (iii) flows with finite, but low $Re_p$, and (iv) arbitrary interpolation and distribution functions, it is limited to unbounded flows. \cite{pakseresht2020} extended this idea to wall-bounded flows by using empirical expressions as well as wall-modified Stokeslet solution. Their approach is applicable to large size particles and extreme anisotropic grids, typically employed in wall-bounded turbulent flows. Test cases performed on velocity of a particle in both parallel and wall-normal motions showed that when the correction schemes, that are developed for unbounded regimes, are employed for correcting particle force in wall-bounded regimes, errors on the same order of magnitude of the uncorrected PP scheme can be obtained. Their model is capable of recovering the undisturbed fluid velocity at any arbitrary wall distance and asymptotically approaches the regular unbounded correction schemes for particles traveling sufficiently away from the no-slip wall.
The above correction schemes by~\citet{esmaily2018,pakseresht2020} remove the self-induced disturbance for each individual particle when correcting the particle forces, keeping the effects of all the other particles in the neighborhood unchanged. Thus, their approaches implicitly account for the effect of neighbors on the individual particle force closures; however, in its present form is limited to $Re_p{<}10$ and tri-linear interpolation on rectilinear grids.

Given the importance of the undisturbed field and the restrictions of the existing models, in this work, a general formulation for estimating the disturbance field created by particles is derived. 
Solution to the disturbance field is obtained using two approaches: (i) direct computation of the disturbance velocity and pressure using the particle reaction forces at computational control volumes, and (ii) an approximate computation of the disturbance velocity field, based on low particle Reynolds number assumption. Direct solution of the disturbance field is easily feasible using the same framework of the Navier-Stokes solver and is applicable to wall-bounded flows, complex geometries and boundary conditions, anisotropic as well as arbitrary, unstructured grids, and a wide range of $Re_p$. The approach is straightforward and can be applied to any numerical formulation for particle-laden flows including Euler-Lagrange as well as Euler-Euler formulations. This direct solution approach does add additional computational cost, but its versatility, simplicity, and accuracy make it an attractive alternative. To reduce the computational cost, and yet keep the same benefits mentioned above, an approximate solution, specifically applicable for low $Re_p$, is also presented. Predictions from these two approaches are compared with existing uncorrected models for motion of a particle in unbounded and wall bounded regimes. Furthermore, the effectiveness of these approaches for a range of grid types (structured or arbitrary shaped unstructured), grid anisotropy, and particle Reynolds numbers is evaluated. 

The rest of the paper is arranged as follows. Section~\ref{sec:formulation} explains the existing issue in the force computations of the standard two-way coupled PP simulations. The mathematical formulations for the direct as well as approximate methods are derived in this section, as well. Section \ref{sec:results} validates both methods on a series of numerical test cases for a particle's motion in unbounded and wall-bounded regimes using various grids and flows parameters. Section \ref{sec:conclusion} concludes the paper with final remarks and summary of the work. 

\section{Mathematical Formulation}
\label{sec:formulation}
In this section, the existing issue in the force computations of the standard two-way coupled PP approaches is first explained. Next, a general framework for correcting this issue is presented followed by a reduced order approximate, but computationally efficient method. For sake of simplicity, pressure-based incompressible fluid flow solvers are used here, however, the proposed framework can be easily extended to any general flow solution techniques including those for compressible flows. Moreover, for the simulations presented in this work, an Euler-Lagrange framework is employed with that in mind that the present methods can be applied to Euler-Euler formulations, as well. 

\subsection{The Issue}
\begin{figure}
    \centering
    \includegraphics[scale=0.5]{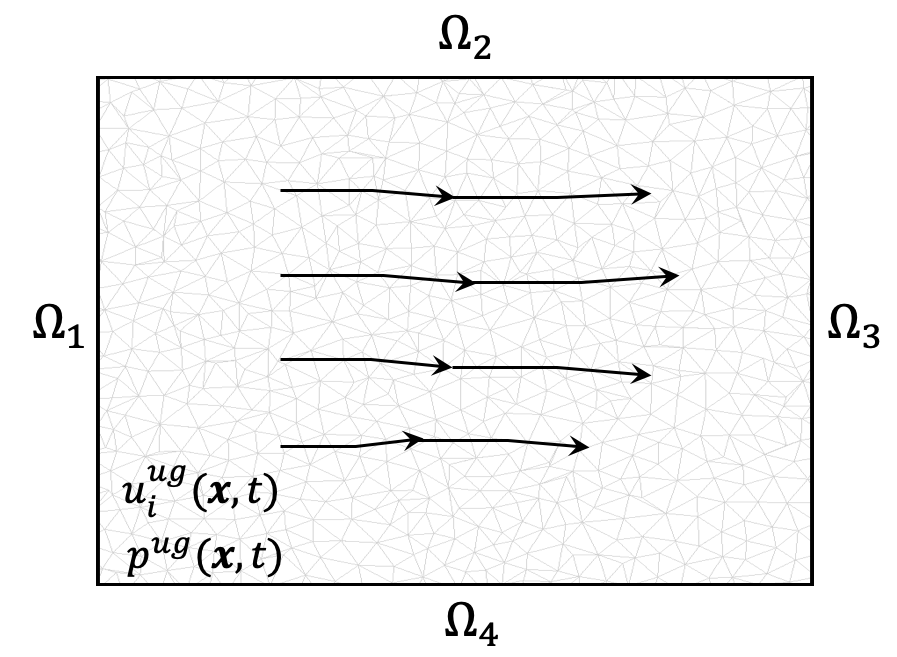}\hspace{0.045\textwidth}
    \includegraphics[scale=0.5]{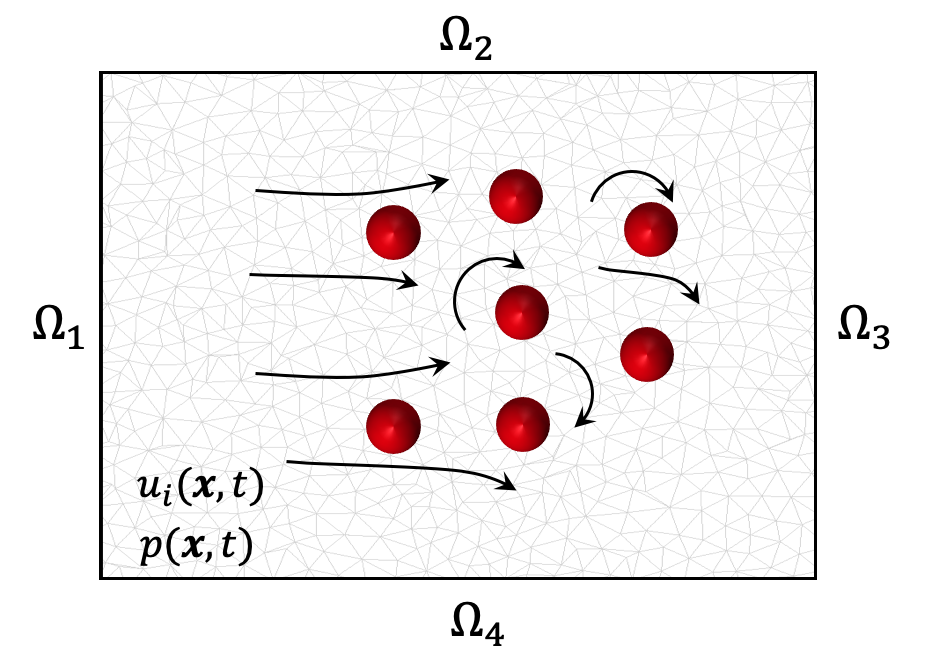}
    \caption{Schematic diagram of flow over particles in a general domain with arbitrary boundary conditions and grids: the globally, undisturbed flow (left) and two-way coupled flow field (right).}
    \label{fig:flow_field}
\end{figure}

Consider a particle-laden fluid flow as shown in Figure~\ref{fig:flow_field}. In a typical point-particle approach in an Euler-Lagrange framework, the particles are assumed subgrid, and their motion is modeled by Newton's second law as,
\begin{eqnarray}
\label{eq:xp}
\frac{d x_{i,p}}{dt} &=& u_{i,p}, \\
\frac{d u_{i,p}}{dt} & = & \left(1 - \frac{\rho_f}{\rho_p}\right) g_i +\frac{1}{\mathcal{V}_p\rho_p}F^t_{i,p},
\label{eq:up}
\end{eqnarray}
\noindent where $x_{i,p}$ and $u_{i,p}$ are particle centroid location and velocity, respectively, $\rho_p$ is the particle density, $\mathcal{V}_p$ is the volume, and $F^t_{i,p}$ represents the total fluid forces acting on each individual particle in the $i$ direction. In the point-particle approach, since particles are assumed subgrid, the forces can not be computed directly and instead are {\it modeled} using the available closures for drag, added mass, history effect \citep{maxey1983}, lift force \citep{saffman1965} and Magnus effect \citep{rubinow1961}, among others. 

In the two-way coupling approach, the effect of the particles on the carrier fluid is modeled through an equal and opposite reaction force of the particles and using an appropriate interpolation kernel, modifying the fluid flow in the vicinity of the particles,
\begin{eqnarray}
\label{eq:moment_2way}
\frac{\partial u_i}{\partial t} + u_j \frac{\partial  u_i}{\partial x_j} &=& -\frac{1}{\rho_f} \frac{\partial p}{\partial x_i} + \nu\frac{\partial^2 u_i}{\partial x_j^2} - \frac{1}{\rho_f}\sum_{p=1}^{N_p} f^t_{i,p}\xi^{\Delta}(\mathbf{x}_{cv}-\mathbf{x}_{p}), \\
\label{eq:cont_2way}
\frac{\partial u_j}{\partial x_j}&=&0,
\end{eqnarray}
\noindent where $\nu$ is the fluid kinematic viscosity, $\rho_f$ is the fluid density, and $f^t_{i,p}$ is the particle force per unit volume in the $i$ direction. $\xi^{\Delta}(\mathbf{x}_{cv}-\mathbf{x}_p)$ is a kernel function to project the particles forces, that lie within the bandwidth ($\Delta$), to the computational cell center at $\mathbf{x}_{cv}$. $N_p$ is the total number of particles that are located within the bandwidth of the projection function. The choice of this projection function is dependent upon the flow under consideration and accuracy needed.

Typical force closure models used in computing the particle motion are based on the fluid flow field that is {\it undisturbed} by the presence of the particle. As an example, the steady state Stokes drag force over a particle with diameter of $d_p$ moving with velocity of $u_{i,p}$ in a fluid with dynamic viscosity of $\mu$ is
\begin{equation}
    F^{Stokes}_{i,drag} = 3\pi\mu d_p\left(u^{un}_{i,@p} - u_{i,p}\right),
   \label{eq:stokes_drag}
\end{equation}
\noindent wherein $u^{un}_{i,@p}$ is the undisturbed fluid velocity at the location of the particle that is not influenced by the presence of the particle under consideration---that is, without the particle self-induced disturbance. However, this undisturbed flow field is not readily available in a two-way coupled simulation as the self-induced disturbance in the fluid flow created by the reaction force of the particle alters the flow velocity and pressure fields.
It should be noted that, the undisturbed fluid flow needed in the closure models for the motion of a particle refers to the velocity and pressure fields in the absence of that particular particle, however, accounts for the disturbance effect created by any of the neighboring particles. 

In the present work, a general formulation is developed to compute the undisturbed flow field that removes the disturbance created by {\it all} particles and is denoted as $u^{ug}_i$ and $p^{ug}$, a globally undisturbed velocity and pressure fields. This approach then allows formulating equations for the disturbance field created by all particles in an Eulerian frame. For a particle under consideration, the undisturbed flow field at the particle location can then be written as,
\begin{equation}
    u_{i,@p}^{un} = u_{i,@p}^{ug} + \sum_{nbr} \delta u_{i,@p}^{nbr}, 
\end{equation}
where $\delta u_{i,@p}^{nbr}$ is the velocity perturbation created by a neighboring particle at the location of the particle $p$ and in its absence. This neighboring effect can be substantial for regimes where inter-particle distance is comparable to the particle size.
The mean inter-particle distance varies as $\phi^{-1/3}$, where $\phi$ is the local particle volume fraction. For example, with $\phi=0.01$, the nearest neighbor distance is about 3.7 times the particle diameter~\citep{akiki2017_jfm}. 
For systems with dilute to moderate volume loadings of $\phi \leq 10^{-3}$, the hydrodynamic inter-particle interactions are insignificant, and thus the effect of neighboring particles on the undisturbed flow field is negligible.  In this regime, $u^{ug}_i \sim u^{un}_i$ and $p^{ug} \sim p^{un}$. In the present work, emphasis is placed on recovering the global, undisturbed flow field ($u^{ug}_i$, $p^{ug}$). The formulation is thus directly applicable to dilute loadings, and can be applied to moderate-to-dense loadings by explicitly incorporating the neighboring particle effects in the future~\citep{moore2019hybrid,seyed-ahmadi_2020}.

To recover the global undisturbed flow field, a general framework is proposed wherein governing equations for the disturbance created by all particles in the flow are first formulated. In the absence of any particles, the fluid flow equations can be written as,
\begin{eqnarray}
\label{eq:moment_und}
\frac{\partial u_i^{ug}}{\partial t} + u_j^{ug} \frac{\partial  u_i^{ug}}{\partial x_j} &=& -\frac{1}{\rho_f}\frac{\partial p^{ug}}{\partial x_i} + \nu\frac{\partial^2 u_i^{ug}}{\partial x_j^2},\\
\frac{\partial u_j^{ug}}{\partial x_j}&=&0.
\label{eq:cont_und}
\end{eqnarray}
The velocity ($u_i^{d}$) and pressure ($p^{d}$) disturbance fields created by all the particles can be obtained by subtracting the disturbed two-way coupled flow field, expressed by Eqs.~\ref{eq:moment_2way}-\ref{eq:cont_2way}, from the undisturbed flow field, given by Eqs.~\ref{eq:moment_und}-\ref{eq:cont_und},
\begin{eqnarray}
\label{eq:mom_disturb}
\frac{\partial u_i^d}{\partial t} + u_j^{d} \frac{\partial  u_i^{ug}}{\partial x_j}+ u_j \frac{\partial  u_i^d}{\partial x_j}  &=& -\frac{1}{\rho_f} \frac{\partial p^d}{\partial x_i} + \nu\frac{\partial^2 u_i^d}{\partial x_j^2} + 
\frac{1}{\rho_f}\sum_{p=1}^{N_p} f^t_{i,p}\xi^{\Delta}(\mathbf{x}_{cv}-\mathbf{x}_{p}) \\
\frac{\partial u_j^d}{\partial x_j}&=&0,
\label{eq:cont_disturb}
\end{eqnarray}
where,
\begin{eqnarray}
\label{eq:vel_und}
u_i^{ug} &=& u_i + u_i^{d},\\
p^{ug} &=& p + p^d.
\label{eq:press_und}
\end{eqnarray}
Here, $u_i$ and $p$ represent the standard two-way coupled velocity and pressure fields, $u_i^d$ and $p^d$ are the disturbance fields created by all particles. The above equation has the boundary condition of $u_i^d{=}0$ far away from the particles. If the particle is near a no-slip wall, the disturbance field also experiences the same no-slip condition of $u_i^d{=}0$, making it a general formulation for any flow configuration, computational approach, and grid type. Solution to the above equations can be obtained by using two different approaches as described below.

\subsection{Direct Method}
In order to solve the equations~\ref{eq:mom_disturb} and \ref{eq:cont_disturb}, the second, nonlinear term on the left hand side of the momentum equation, $u^d_j\partial u^{ug}_i/\partial x_j$, requires additional closure between the undisturbed and disturbance fields. However, this term can be safely neglected in comparison to the third term by hypothesizing the following,
\begin{equation}
\frac{\partial  u_i^{ug}}{\partial x_j} \ll \frac{\partial  u_i^d}{\partial x_j},
\end{equation}
\noindent which is approximately valid as the velocity gradient caused by the particle force in the disturbance field is conjectured to be greater than that of the undisturbed field. Although this assumption is verified by the small errors for the studied cases reported in section \ref{sec:results}, further investigations for cases with inherently large velocity gradient in the undisturbed field is left for future investigations. Knowing the particle forces, the disturbance field can then be obtained by directly solving the following equations,
\begin{eqnarray}
\label{eq:moment_dist}
\frac{\partial u_i^d}{\partial t} + u_j \frac{\partial  u_i^d}{\partial x_j} &=& -\frac{1}{\rho_f} \frac{\partial p^d}{\partial x_i} + \nu\frac{\partial^2 u_i^d}{\partial x_j^2} + \frac{1}{\rho_f}\sum_{p=1}^{N_p} f^t_{i,p}\xi^{\Delta}(\mathbf{x}_{cv}-\mathbf{x}_{p}), \\
\frac{\partial u_j^d}{\partial x_j}&=&0.
\label{eq:cont_dist}
\end{eqnarray}
Note that the nonlinear, advective term contains the disturbance velocity ($u_i^d$) and the two-way, coupled velocity ($u_j$). The latter is readily available in a two-way coupled simulation.

The same computational algorithm employed for the solution of the main two-way coupled flow field (Eqs. \ref{eq:moment_2way}-\ref{eq:cont_2way}) can be utilized to compute the disturbance field (equations above), which involves solution of the disturbance momentum equations, and projection of the divergence-free disturbance condition using a solution of a Poisson equation for the disturbance pressure. Direct solution of the disturbance field is then used to recover the undisturbed flow field from Eqs~\ref{eq:vel_und}-\ref{eq:press_und} and to accurately compute the fluid forces acting on the particles. Compared to the existing correction schemes, the direct method benefits from many advantages as explained below:

\begin{itemize}
\item Direct solution is easily feasible using the same framework employed for solving the Navier-Stokes equations of the two-way coupled field.

\item Wall-bounded flows, complex geometries, and arbitrary boundary conditions can be automatically accounted for in the solution of the disturbance field. 
\item Unlike the majority of the existing models, the direct method is capable of handling a wide range of $Re_p$. 
\item The disturbance velocity and pressure fields (and thus their gradients) are available, thus the common force closures for drag, lift, history effect, as well as pressure gradient forces can be accurately computed. 
\item The formulation is free of any tuning or empirical expressions, typically used for specific grid configurations, and can be applied to both structured and arbitrary shaped, hybrid unstructured grids with any grid aspect ratio. 
\item The formulation is free of any dependency on the interpolation and projection functions employed in the two-way coupled simulations. 
\item The disturbance field is computed regardless of size of particles, hence it is adaptable for flows with any arbitrary size particles, particularly those with particles larger than grid. 
\end{itemize}

The main drawback of this approach is the additional computational cost for solving the disturbance field that requires full solution of the momentum as well as continuity equations, which makes the computations as nearly twice as expensive. The additional cost can still be tolerable for direct numerical or large eddy simulations, as the approach is much more affordable than particle-resolved methods. However, for a faster computation, an approximate method is introduced in the following part, which is shown to be reasonably accurate as compared to the direct method while being significantly more cost efficient. 

\subsection{Approximate Method}
\label{sec:approximate}
In this part, an approximate solution of the disturbance field is proposed that is derived based on low particle Reynolds number assumption. In the limit of creeping flow ($Re_p{<}0.1$), the inertial terms on the left hand side of Eq. \ref{eq:moment_dist} are dropped and the simplified momentum and continuity equations then become,
\begin{eqnarray}
\label{eq:moment_appx}
\frac{\partial u_i^d}{\partial t} &=& -\frac{1}{\rho_f} \frac{\partial p^d}{\partial x_i} + \nu\frac{\partial^2 u_i^d}{\partial x_j^2} + \frac{1}{\rho_f}\sum_{p=1}^{N_p} f^t_{i,p}\xi^{\Delta}(\mathbf{x}_{cv}-\mathbf{x}_{p}), \\
\frac{\partial u_j^d}{\partial x_j}&=&0.
\label{eq:cont_appx}
\end{eqnarray}
In order to further simplify the equations above, we conjecture that the fluid response to the particle force, is approximately analogous to the flow that would be generated by the particle in the real physics of the problem. This is in fact the main assumption employed in the two-way coupled point-particle approach wherein it is assumed that the particle force can approximately produce the same flow as the particle would do in the reality. Motivated by this analogy and in the limit of steady state and $Re_p{<}0.1$, we recall the Stokes solution that is the flow created around the actual particle. In the Stokes regime, the drag on the particle that experiences slip velocity of $u^{rel}_p$, consists of two terms (i) pressure and (ii) viscous forces as, 

\begin{equation}
F_{\rm drag}^{\rm Stokes} = \underbrace{\pi \mu d_p u^{rel}_p}_{\rm Pressure ~ force} + \underbrace{2\pi \mu d_p u^{rel}_p}_{\rm Viscous ~force}.
\end{equation}

\noindent For low particle Reynolds numbers, the expression for these two forces are identical, with viscous force being twice greater than the pressure force. Motivated by this, one can model the contribution of the pressure drag through an effective viscosity and rewrite the Stokes drag force as,

\begin{equation}
F_{i,\rm drag}^{\rm Stokes} = 2\pi \mu_{eff} d_p u^{rel}_p; \quad \mu_{eff} = K_{\nu}\mu;~~~ K_{\nu} = 1.5.
\end{equation}

Rewriting the Stokes drag in this form facilitates the approximation that the effect of the pressure gradient term in Eq.\ref{eq:moment_appx} can be modeled through an equivalent viscous term with an effective viscosity of $K_{\nu}{=}1.5$ to match the net fluid force in the Stokes limit. It should be noted that the continuity constraint is already embedded in the Stokes solution from which the Stokes drag is obtained. Therefore, it is conjectured that the introduced correction factor will implicitly provide a velocity field that approximately satisfies the continuity equation.  Since, the pressure term is no longer needed, the continuity constraint in Eq.\ref{eq:cont_appx} is unnecessary and is only satisfied approximately. Using this approximate method, the disturbance field due to the particles forces, can be computed by solving only the momentum equation in each direction with viscous stresses and a modified viscosity through the introduced correction factor of $K_{\nu}{=}1.5$. The approximate equation then becomes,
\begin{eqnarray}
\frac{\partial u_i^d}{\partial t} &=& K_{\nu}\nu\frac{\partial^2 u_i^d}{\partial x_j^2} + \frac{1}{\rho_f}\sum_{p=1}^{N_p} f^t_{i,p}\xi^{\Delta}(\mathbf{x}_{cv}-\mathbf{x}_p).
\label{eq:approximate}
\end{eqnarray}
It is worth mentioning that the correction factor $K_{\nu}$ can be Reynolds number dependent. With increase in $Re_p$, the contribution of the pressure drag to the net drag is bound to increase~\citep{white2006viscous}, and the value of $K_{\nu}$ can potentially be changed. For the present study, however, the value is kept fixed and equal to 1.5, even for higher $Re_p$, and further adjustment for larger $Re_p$ cases are left for future. 
Similar to the direct method, the equation above is solved in the same Eulerian frame that is used for solving the two-way coupled flow field equations.
This captures the resultant disturbance field that is caused by all particles in the flow field. For dilute loadings, wherein the disturbance of each particle is isolated from that of the neighbouring particles, this model perfectly captures the self-induced disturbance of each particle required in force closures. However, for dense loadings, the neighbouring effect will be removed and such an effect should be added explicitly using the recently developed models by \citet{moore2019hybrid,seyed-ahmadi_2020}. 
Depending on the application of interest, both unbounded and wall-bounded regimes can be handled by this method since the boundary conditions are directly enforced for solving the equation above. Finally, owing to the linear, unsteady diffusion like equation with a source term, its solution is considerably faster than the direct method.
Note that Eq.~\ref{eq:approximate} is general and directly applicable to any arbitrary grid. Concerning the applicability of this method for $Re_p{>}0.1$, it is shown later (section \ref{sec:results}) that despite the fact that this method is constructed upon the assumption of small $Re_p$, it can reduce the errors for a wider range of particle Reynolds numbers of $Re_p{\leq}100$.

\subsection{Numerical Algorithm}
The procedure in the present disturbance-corrected point-particle (DCPP) framework is similar to the standard uncorrected point-particle approach with an additional step for recovering the undisturbed flow field. For the computations of the present methods, two sets of parameters and equations, corresponding to the disturbance as well as two-way coupled disturbed flow fields, are solved separately yet on similar computational domains and identical boundary conditions. Note that any interpolation and projection functions that are used for computations of the two-way coupled flow field should be used for the computation of the disturbance field, to ensure that the predicted disturbance is consistent with the one that particles actually sample in the disturbed two-way coupled flow field. Knowing the computed disturbance field, $u^d_{i}$, and particles velocity, $u_{i,p}$, from the previous time step, the following procedure is employed. \\

\noindent 1. Solve Eqs. \ref{eq:moment_2way} and \ref{eq:cont_2way} for the two-way coupled field to update the fluid velocity, $u_i$, and pressure, $p$, due to presence of particles. Note that this is the standard step in the uncorrected PP approaches. \\ 
2. Knowing the disturbance field available from previous time step, and the updated disturbed flow field from step 1, recover the undisturbed fluid velocity, $u^{ug}_i$, and pressure, $p^{ug}$, fields at the location of particles using Eqs. \ref{eq:vel_und} and \ref{eq:press_und}.\\
3. Use the undisturbed field to compute the net fluid forces acting on each particle, $F^t_{i,p}$.\\
4. Update the velocity and location of each individual particle using Eqs. \ref{eq:xp} and \ref{eq:up}.\\ 
5. Knowing the particles reaction forces from step 3, compute the disturbance field by solving either the direct method (Eqs. \ref{eq:moment_und} and \ref{eq:cont_und}) or the approximate method (Eq.\ref{eq:approximate}). 

Present work is based on an energy-conserving scheme for unstructured, arbitrarily shaped grid elements based on fractional time-stepping on a colocated mesh~\citep{mahesh2004numerical}.
The velocity and pressure are stored at the centroids of the volumes. The cell-centered velocities are advanced in a predictor step, the predicted velocities are interpolated to the faces and then 
projected. Projection yields the pressure potential at the cell-centers, and its gradient is used to correct the cell and face-normal velocities, using an area weighted least-squares minimization technique~\citep{mahesh2004numerical}. Details of the algorithm on arbitrary shaped unstructured grids for particle-laden flows are given in~\cite{shams2011numerical} and a brief description is presented in Appendix A for completeness. The same algorithm is used for the disturbance field in the direct and approximate methods.

\section{Results}
\label{sec:results}
In this section, the performance of the direct as well as approximate methods on recovering the undisturbed flow field and correcting the PP approach is verified in various scenarios. A single particle settling under gravity in an unbounded regime is investigated first. Settling velocity of the particle moving parallel and normal to a no-slip wall is performed next. As the final test case, the unsteady motion of a single particle in an oscillatory field is examined, as well. 
For simplicity, drag force as the only fluid force acting on the particle is employed while other fluid forces such as lift, added mass, pressure gradient, and history effect are neglected. For each set, various grid configurations including isotropic and anisotropic rectilinear grids as well as tetrahedral unstructured grid are used to assess the accuracy of the present models on arbitrary shaped grids. A range of $0.1{\leq}Re_p{\leq}100$ is performed to evaluate the models for a wide range of scenarios that may happen in particle-laden flows. The grid resolution of $128^3$ was chosen for all cases (with close proximity to this resolution for the unstructured grid) as it was found to be sufficient to produce the grid-independent results. 

Three non-dimensional parameters are used to setup the cases: (i) particle-to-gird size ratio, $\Lambda$, (ii) particle Stokes number, $St$, and (iii) particle Reynolds number, $Re_p$. The first dimensionless parameter, $\Lambda$, is defined as 

\begin{equation}
    \Lambda = \frac{d_p}{d_c},
\end{equation}

\noindent where $d_p$ and $d_c$ are the particle diameter and the characteristic length of the grid, respectively. For rectilinear grids, $d_c$ can become a vector with three components each of which corresponding to the size of the grid in that direction, $a_i$, hence three components for $\Lambda$, as well. However, for unstructured grids, finding an equivalent grid size for each direction is not trivial, therefore, a unified $d_c$ based on the diameter of a sphere that has the equivalent volume of the grid, $d_c{=}\sqrt[3]{6{\mathcal V}_{cv}/\pi}$, is defined. Particle Stokes number is defined as,

\begin{equation}
    St=\frac{\tau_p}{\tau_f},
    \label{eq:st}
\end{equation}

\noindent where, 

\begin{equation}
    \tau_p = \frac{\rho_pd^2_p}{18\mu},
    \label{eq:tau_p}
\end{equation}

\noindent and,
\begin{equation}
    \tau_f = \frac{\min\left(d_c\right)^2}{\nu},
\end{equation}

\noindent are the respective particle relaxation time and fluid time scale in the Stokesian regime. The particle Reynolds number in this regime is also defined as,

\begin{equation}
    Re^{Stk}_p = \frac{|\mathbf{u}^{Stk}_{s}|d_p}{\nu},
     \label{eq:Rep_Stk}
\end{equation}
\noindent where, 
\begin{equation}
    \mathbf{u}^{Stk}_{s}=\left(1-\frac{\rho_f}{\rho_p} \right)\tau_p\mathbf{g},
    \label{eq:u_set_stokes}
\end{equation}
\noindent is the particle settling velocity under gravity vector of $\mathbf{g}$. It is imperative to mention that for the studied cases with $Re^{Stk}_p{>}0.1$ or those in wall-bounded regime, the particle's drag coefficient varies from that of the Stokes flow, so does the particle settling velocity, thus the actual particle Reynolds number, denoted by $Re_p$, differs from Eq. \ref{eq:Rep_Stk}. For each of those cases, the proper expression is provided, separately. 

The fluid velocity at the particle's location, required for the drag force computation, is interpolated using a three-point delta function with compact support that uses the nearest neighbors of a control volumes~\citep{roma1999adaptive}. For control volumes with resolution of 
$\Delta{=}\sqrt[3]{{\mathcal V}_{cv}}$, the interpolation stencil utilizes only three points in one dimension and thus is easiest to implement:
\begin{equation}
\label{eq:delta}
\xi^{\Delta}(\mathbf{x}_{cv}-\mathbf{x}_p) =  \left\{ \begin{array}{ll}
			{1 \over 6} (5-3|r| - \sqrt{-3(1-|r|)^2+1}),& ~~~~0.5\leq |r| \leq 1.5, r=|\mathbf{x}_{cv}-\mathbf{x}_p|/{\Delta} \\
			{1 \over 3} (1+\sqrt{-3r^2+1}),& ~~~~|r| \leq 0.5 \\
			0. & ~~~~{\rm otherwise}
			\end{array}
			\right.	
\end{equation}

Given the force balance acting over the particle, it is advanced using a first order Euler approximation to solve Eqs. \ref{eq:xp} and \ref{eq:up}. Concerning the two-way coupled simulations, the particle reaction force is exerted to the nearby fluid control volumes using identical function as expressed above. 
Although a simple three-point delta function is used for Euler-Lagrange interpolation and projection purposes, the present methods can be easily adopted for any other functions. 
For the computations, we correct the PP approach using both direct and approximate methods and compare their results to those of the uncorrected PP as well as the corresponding reference for each part. The reference is obtained based on the one-way coupled simulations wherein the fluid phase remains undisturbed, and drag force and particle motion are accurately computed.

\subsection{Settling in an unbounded quiescent fluid}
In this part, settling velocity of a single particle in an unbounded quiescent fluid is performed. A particle that is initially at rest settles under a gravity vector and experiences drag force, only. The particle equation of motion then becomes, 

\begin{equation}
\frac{du_{i,p}}{dt} = \left(1-\frac{\rho_f}{\rho_p} \right)g_i - \frac{f}{\tau_p} (u_{i,p}-u_{i,@p}^{ug}),
\label{eq:dup_dt}
\end{equation}

\noindent where $u_{i,@p}^{ug}$ is the interpolated fluid velocity at the location of the particle that is erroneously nonzero in the uncorrected two-way coupled simulations, owing to the disturbance created by the particle in the nearby computational cells. The direct as well as approximate methods, however, predict and remove this velocity as in the real physics of the problem (and one-way coupled simulations) this velocity is zero. In general, the factor of $f$ can correspond to any adjustment to the Stokes drag to account for different effects. In this part, it follows the Schiller-Naumman adjustment factor \citep{clift}, as expressed below, to account for the finite Reynolds number effect of the particle on the Stokes drag in unbounded regime,

\begin{equation}
    f = 1 + 0.15Re_p^{0.687} ~ ; ~ Re_p = \frac{|\mathbf{u}_p|d_p}{\nu}.
    \label{eq:schiller-Naumman}
\end{equation}

\noindent Such an adjustment results in an effective particle relaxation time, $\tau^{e}_p$, as,

\begin{equation}
    \tau^{e}_p = \frac{\tau_p}{f}.
    \label{eq:taup_eff}
\end{equation}

Following~\cite{horwitz2016}, gravity vector of $\mathbf{g}{=}[1,(1+\sqrt{5})/2,\exp(1)]/|\mathbf{g}|$ is chosen so that particle sweeps through different positions among its adjacent computational cells, ensuring that the present models are capable of handling any arbitrary positioning of the particle. The time step, $\Delta t$, for the computations is also chosen as, 

\begin{equation}
    \Delta t =  {\rm min}\left(0.03\tau_f,0.03\tau^{e}_p,0.3\tau_{cfl}\right),
    \label{eq:delta_t}
\end{equation}

\noindent with $\tau_{cfl}$ being the time scale associated with the fluid advection in high particle Reynolds number cases based on Courant-Friedrichs-Lewy (CFL) condition less than one for time-accurate solutions. Here, a maximum CFL of $0.3$ is assumed.

Accuracy of each model is evaluated in terms of predicting the particle velocity in comparison with the reference. The particle velocity as a function of time, $\mathbf{u}_p(t)$, is decomposed into two components (i) parallel, $\mathbf{u}^{||}_p$, and (ii) normal, $\mathbf{u}^{\perp}_p$ to the reference settling velocity (terminal velocity) of $\mathbf{u}_s$, and are obtained respectively as,

\begin{equation}
    \mathbf{u}^{||}_p(t) = \frac{\mathbf{u}_s \cdot \mathbf{u}_p(t)}{|\mathbf{u}_s|^2}\mathbf{u}_s,
    \label{eq:up_parallel}
\end{equation}

\noindent and 

\begin{equation}
    \mathbf{u}^{\perp}_p(t) = \mathbf{u}_p(t) - \mathbf{u}^{||}_p(t).
    \label{eq:up_perp}
\end{equation}

\noindent The errors in these two velocity components as well as the total particle velocity in a time average manner, denoted by $\overline{()}$, are then calculated using the respective following metrics,
\begin{equation}
e^{\parallel} = \frac{\overline{\mathbf{u}^{\parallel}_p(t).\mathbf{u}_s}}{|\mathbf{u}_s|^2} -1;
\label{eqn:paral_error}
\end{equation}

\begin{equation}
e^{\perp} = \frac{\overline{|\mathbf{u}^{\perp}_p(t)|}}{|\mathbf{u}_s|};
\label{eqn:perp_error}
\end{equation}

\begin{equation}
e = \frac{\overline{|\mathbf{u}_p(t)-\mathbf{u}_s|}}{|\mathbf{u}_s|}.
\label{eqn:perp_error1}
\end{equation}

\begin{figure}
    \centering
    \includegraphics[scale=0.55]{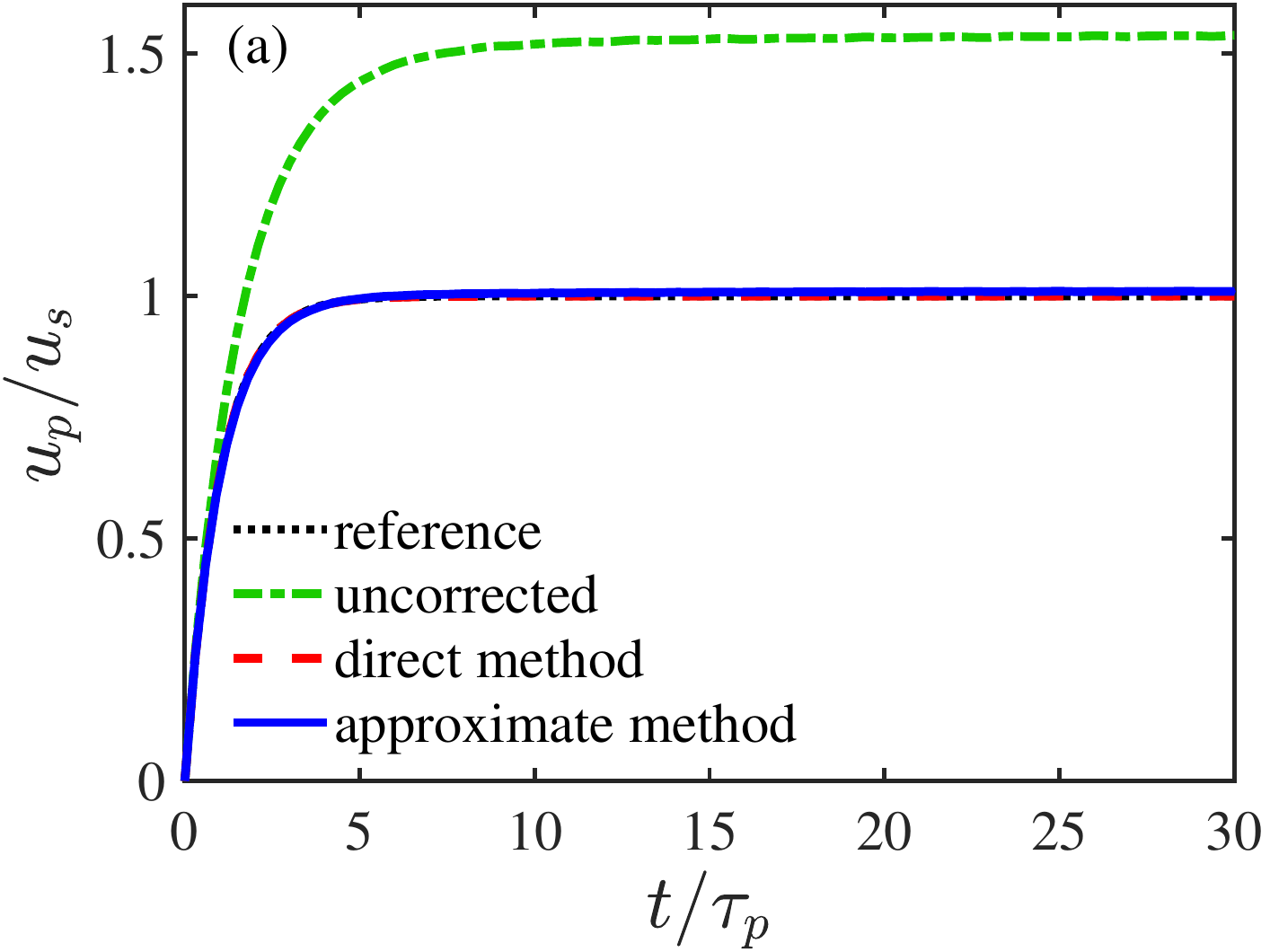}
    \includegraphics[scale=0.55]{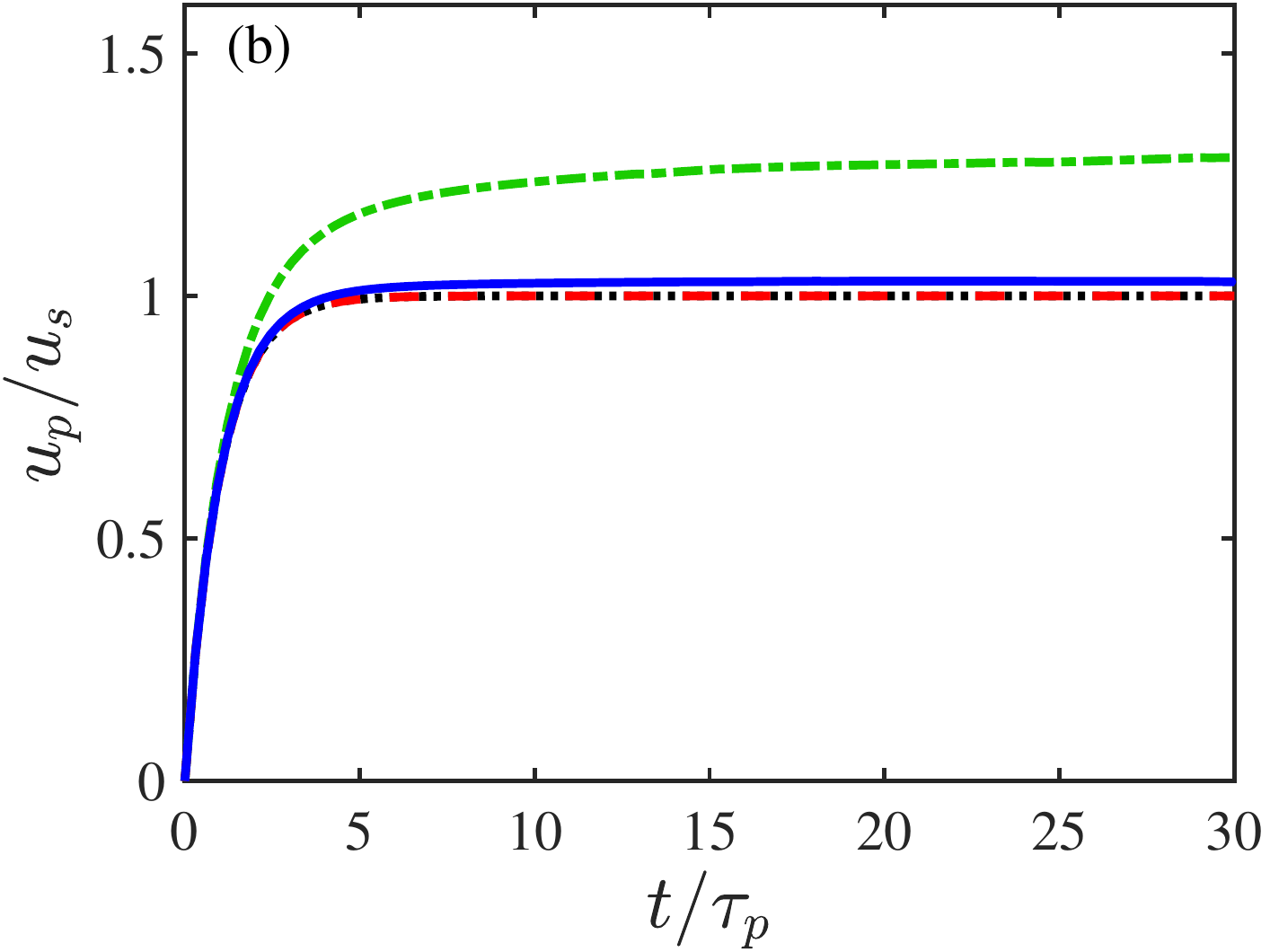}\vspace{0.06\textwidth}
    \includegraphics[scale=0.55]{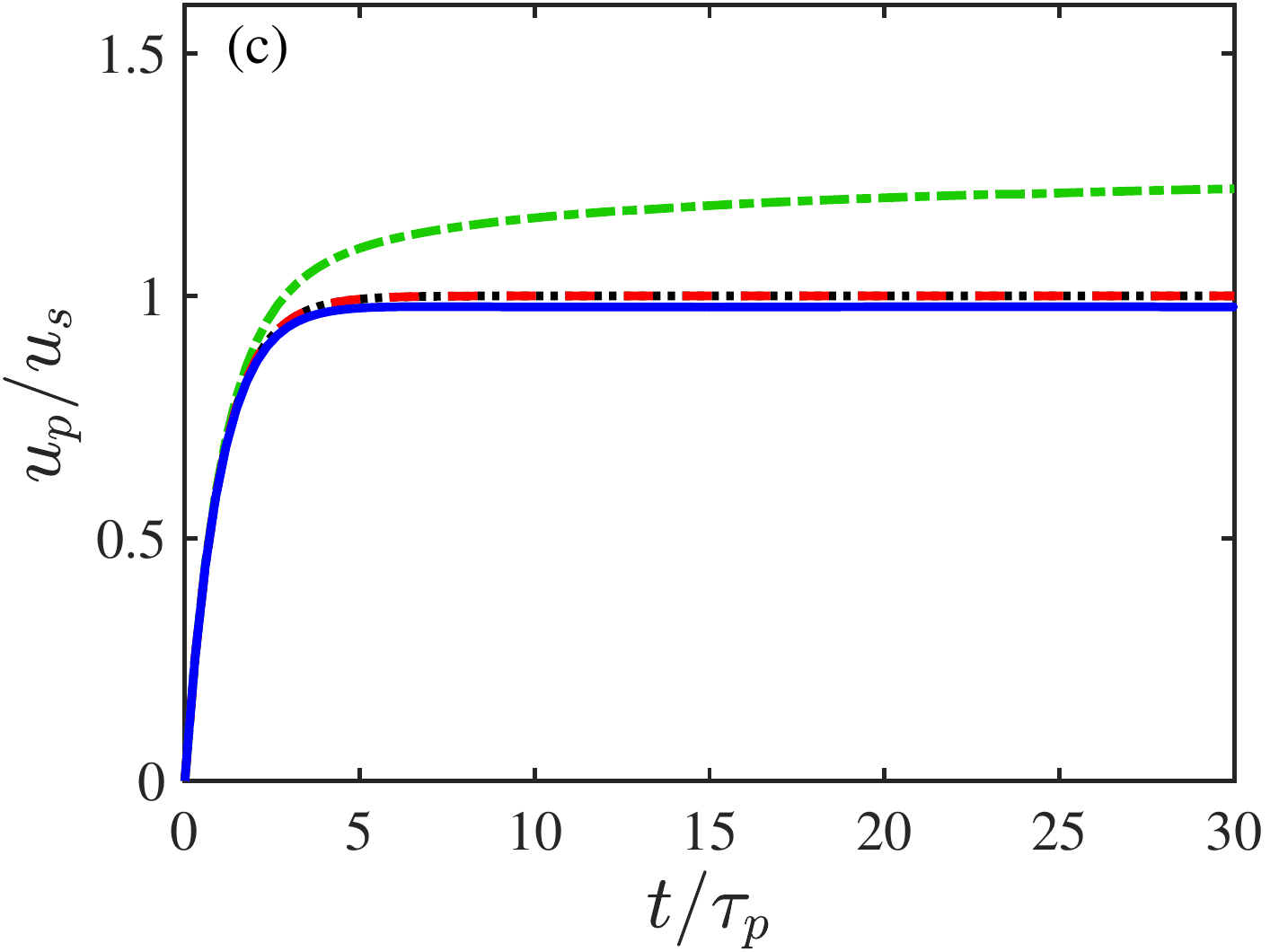}
    \includegraphics[scale=0.55]{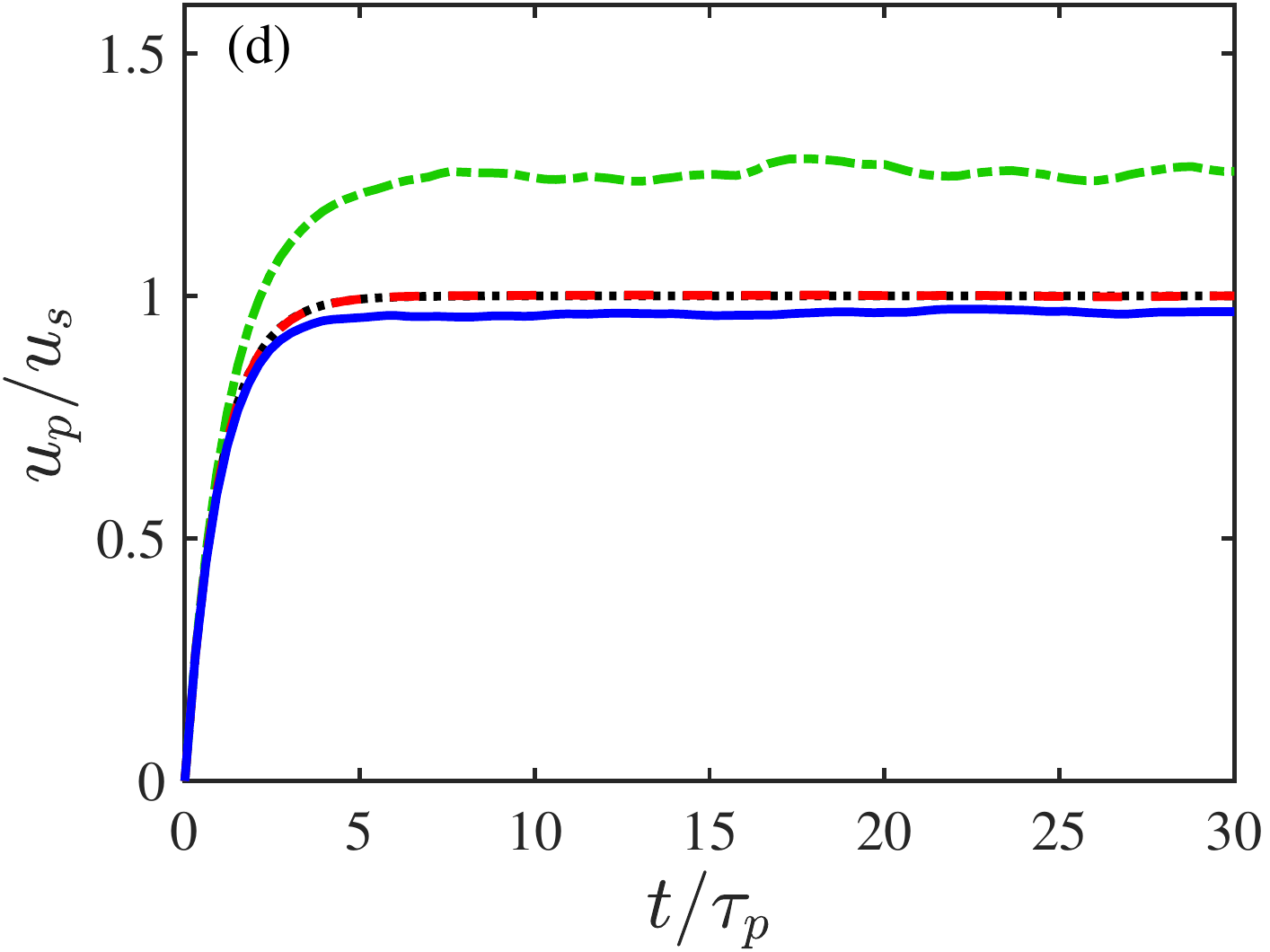}
    \caption{Temporal evolution of particle velocity settling in an unbounded, quiescent flow at $Re_p{=}0.1$ and $St{=}10$, is shown. Plotted are four different cases from Tab.\ref{tab:unbounded_low_re}: (a) isotropic rectilinear grid (case U1), (b) anisotropic rectilinear grid with $\mathbf{\Lambda}{=}[4.0,2.0,2.0]$ (case U5), (c) anisotropic rectilinear grid with $\mathbf{\Lambda}{=}[0.3,6.0,0.6]$ (case U6) and (d) tetrahedral unstructured grid with $\Lambda{=}1.0$ (case U8). Predictions of the approximate method (solid blue), direct method (dashed red), and uncorrected scheme (dash-dotted green) are compared against the reference (dotted black). The particle relaxation time, $\tau_p$, and settling velocity expressed by Eqs. \ref{eq:tau_p} and \ref{eq:u_set_stokes}, respectively, are used for normalizing the results.
    } 
    \label{fig:unbounded_lowRe}
\end{figure}

Table \ref{tab:unbounded_low_re} lists these errors for a particle with $Re_p{=}0.1$ settling over various grid configurations such as isotropic rectilinear grid, anisotropic rectilinear grid, and tetrahedral unstructured grid. For each case, the errors obtained by the corrected PP approach using the direct as well as approximate methods are compared against those of the uncorrected PP scheme.
As explained before, the $\Lambda$ parameter for rectilinear grids is a vector that has three components for the particle-to-grid size ratio in each direction. However, for the unstructured grids, this parameter becomes only a scalar that is obtained based on the size of the particle and the average $d_c$ over the entire grid. 
It is observed that the errors associated with the uncorrected PP approach depend on $\Lambda$ with bigger particles producing stronger disturbances in the background flow. As an example, particle in case U2 produces five times larger errors compared to that of case U1 that has a particle five times smaller. Similar comparison is observed between cases U8 and U9. Concerning the effect of particle Stokes number, results based on two different $St{=}3.0$ and $St{=}10$ (e.g., case U1 and U3, respectively) show small dependency of the errors on this parameter, consistent with the preceding works. When the standard PP approach is corrected using the present methods, however, significant error reduction is observed with nearly zero errors for the direct method across the board. Although the approximate model yields slightly larger errors compared to the direct model, such as those in U2 and U9, the overall errors are still an order of magnitude lower than the uncorrected scheme. The affordability of the approximate method makes it an attractive scheme for recovering the undisturbed flow field, given the fact that the difference in the error reduction between these two methods is still insignificant.  
The errors for approximate method in this case are also comparable to those reported by~\citet{pakseresht2020}. Figure \ref{fig:unbounded_lowRe} qualitatively illustrates the performance of these models in predicting the time-dependent particle velocity for different grid resolutions. The particle relaxation time and settling velocity expressed by Eqs. \ref{eq:tau_p} and \ref{eq:u_set_stokes}, respectively, are used for normalizing the results. 

\begin{figure}
    \centering
    \includegraphics[trim={0.8inch 3.7inch 0.1inch 2.0inch},clip,scale=0.72]{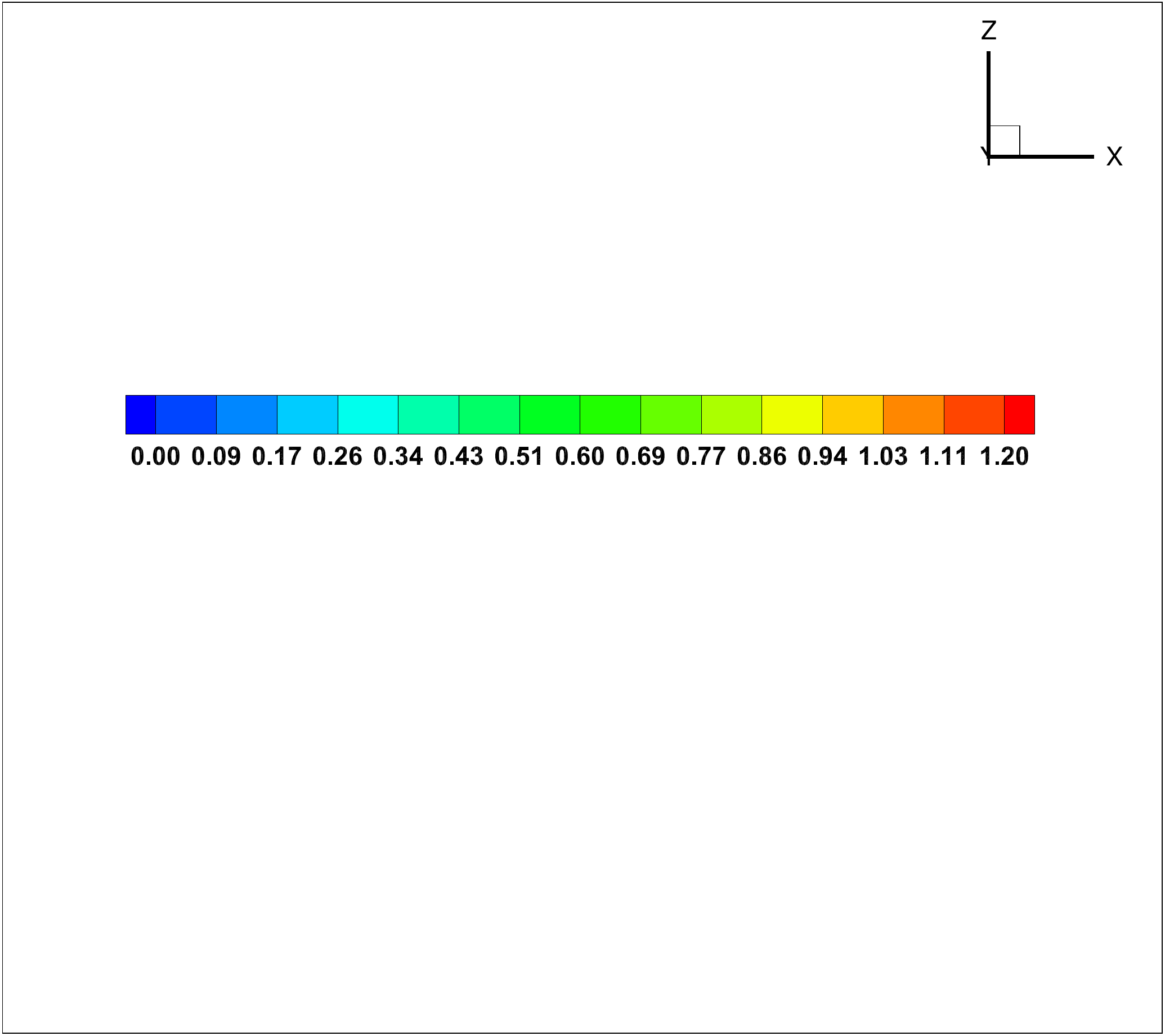}\\
    \includegraphics[trim={2.67inch 2.16inch 2.68inch 2.15inch},clip,scale=0.4]{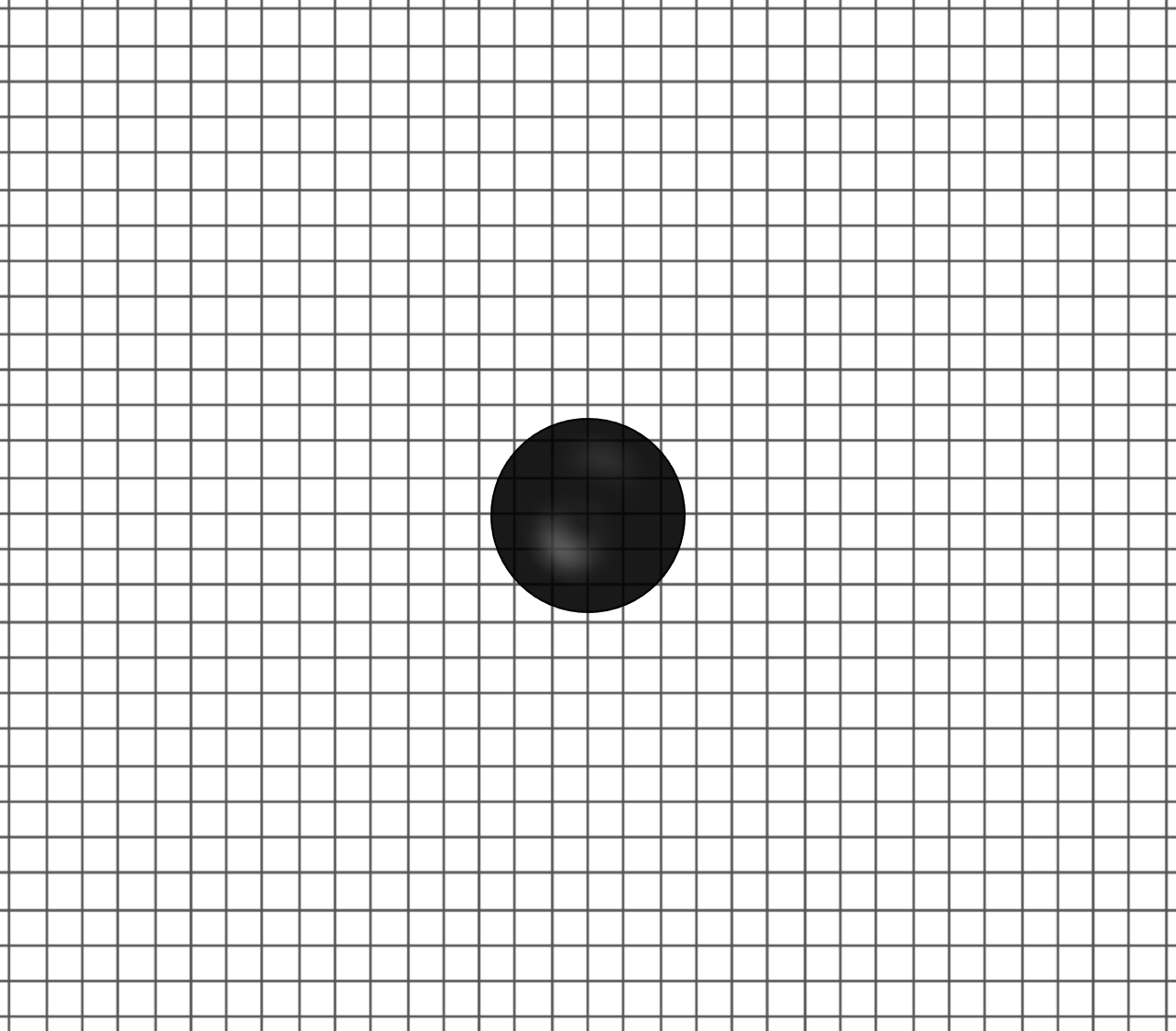}
    \includegraphics[trim={2.6inch 2.05inch 2.5inch 1.8inch},clip,scale=0.47]{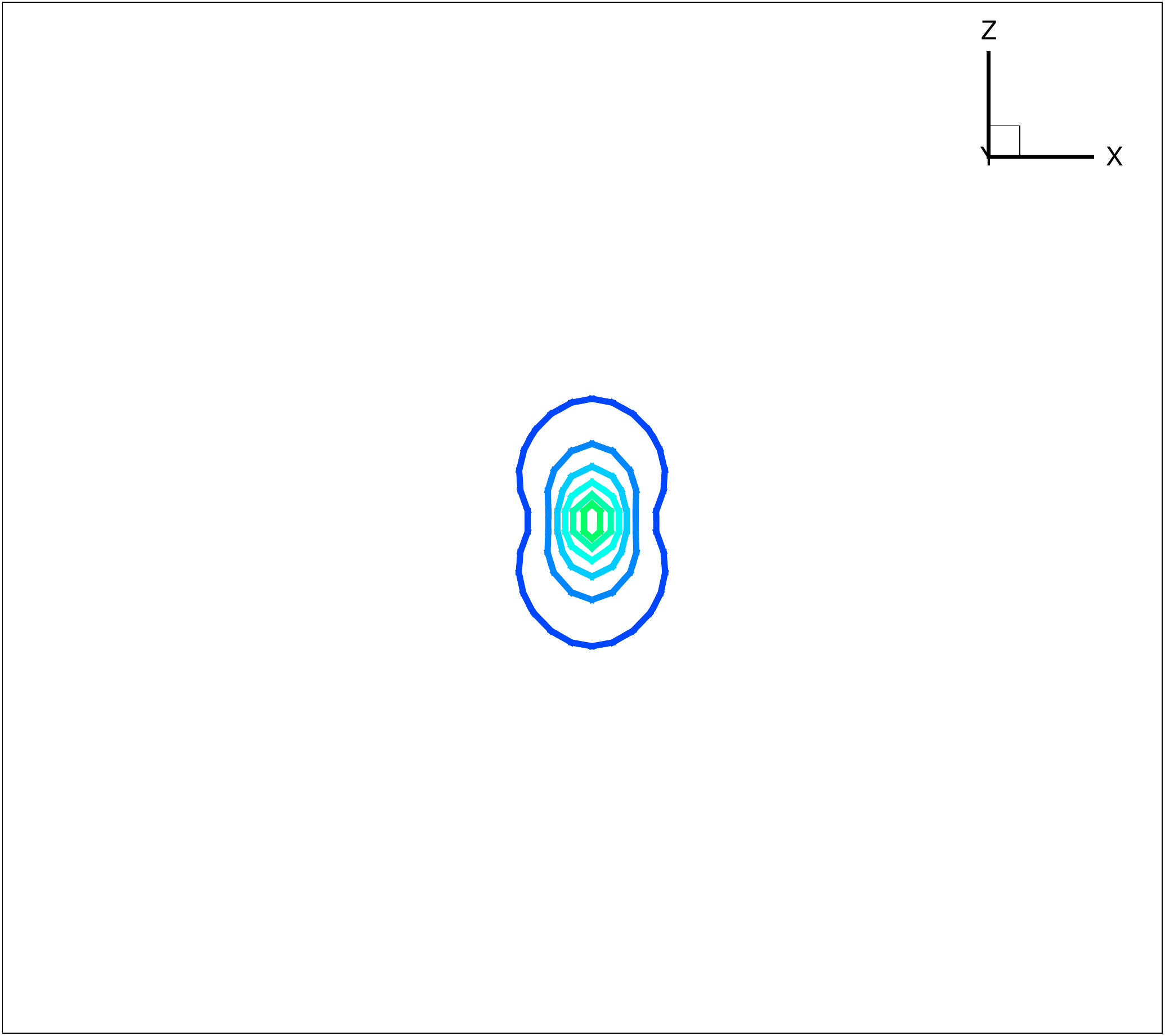}
    \includegraphics[trim={2.8inch 2.05inch 2.2inch 1.8inch},clip,scale=0.47]{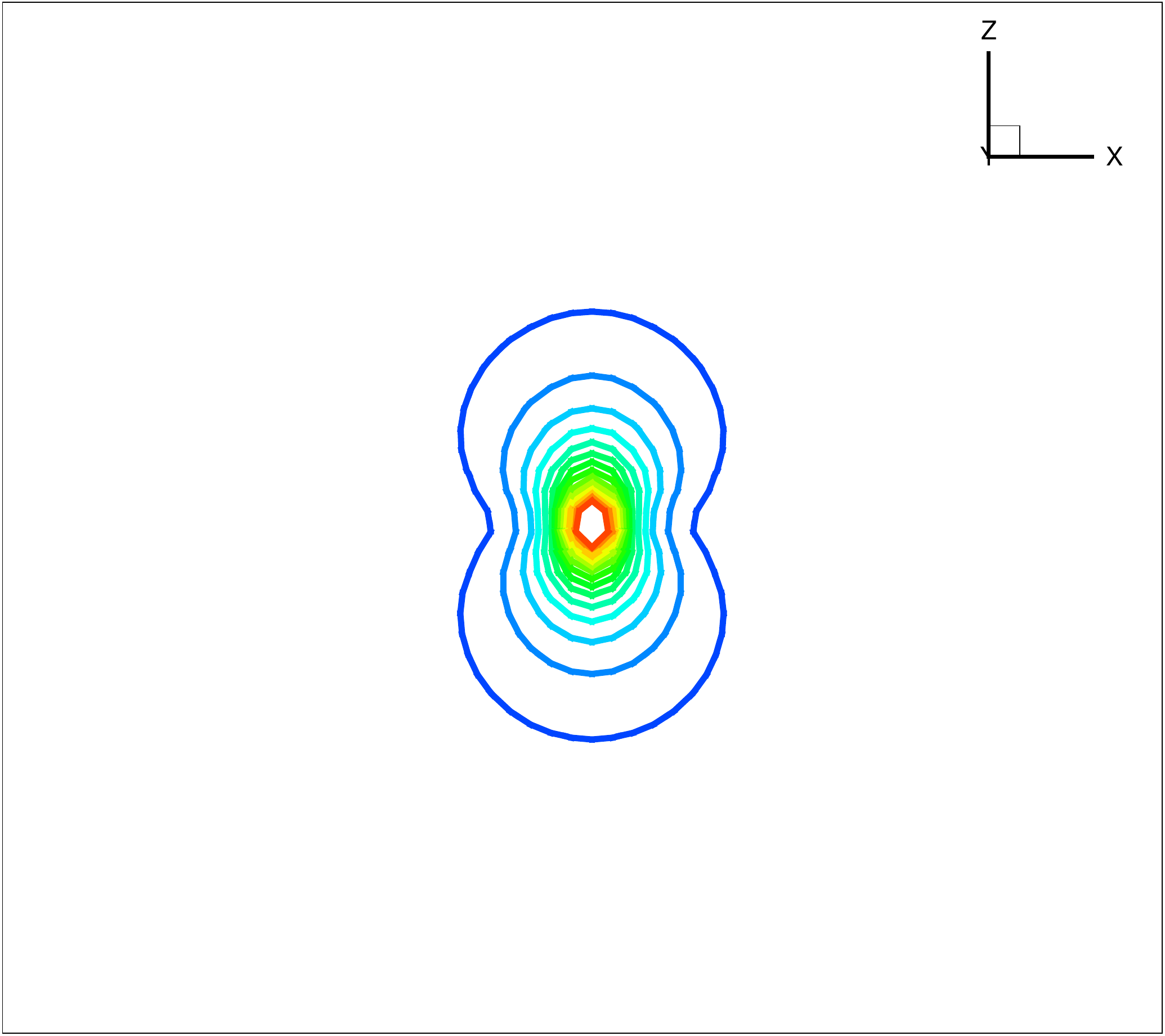}
    \includegraphics[trim={2.8inch 2.05inch 2.2inch 1.8inch},clip,scale=0.47]{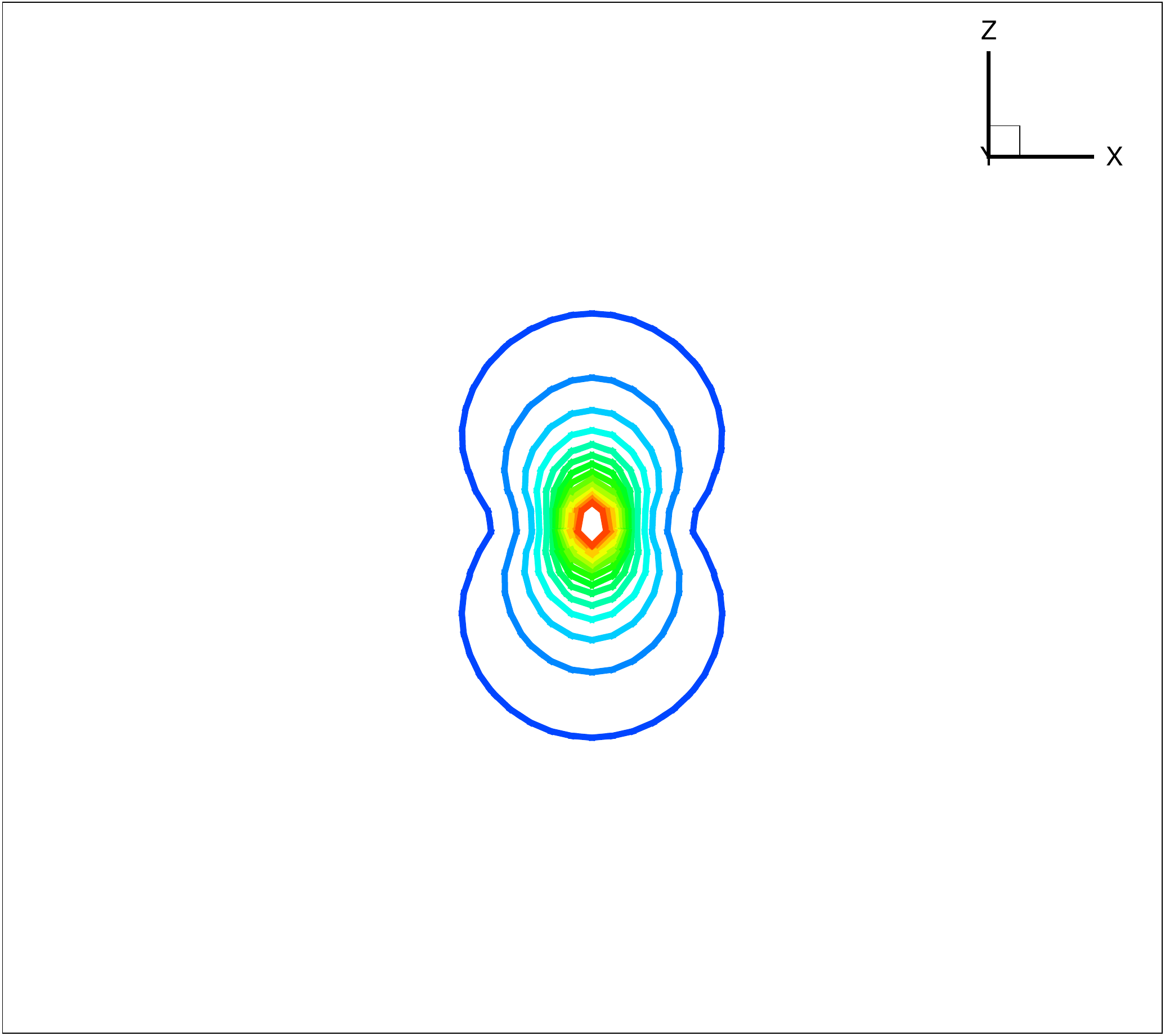}
    
    \vspace{0.04\textwidth}
    \includegraphics[trim={2.65inch 2.2inch 2.4inch 2.0inch},clip,scale=0.4]{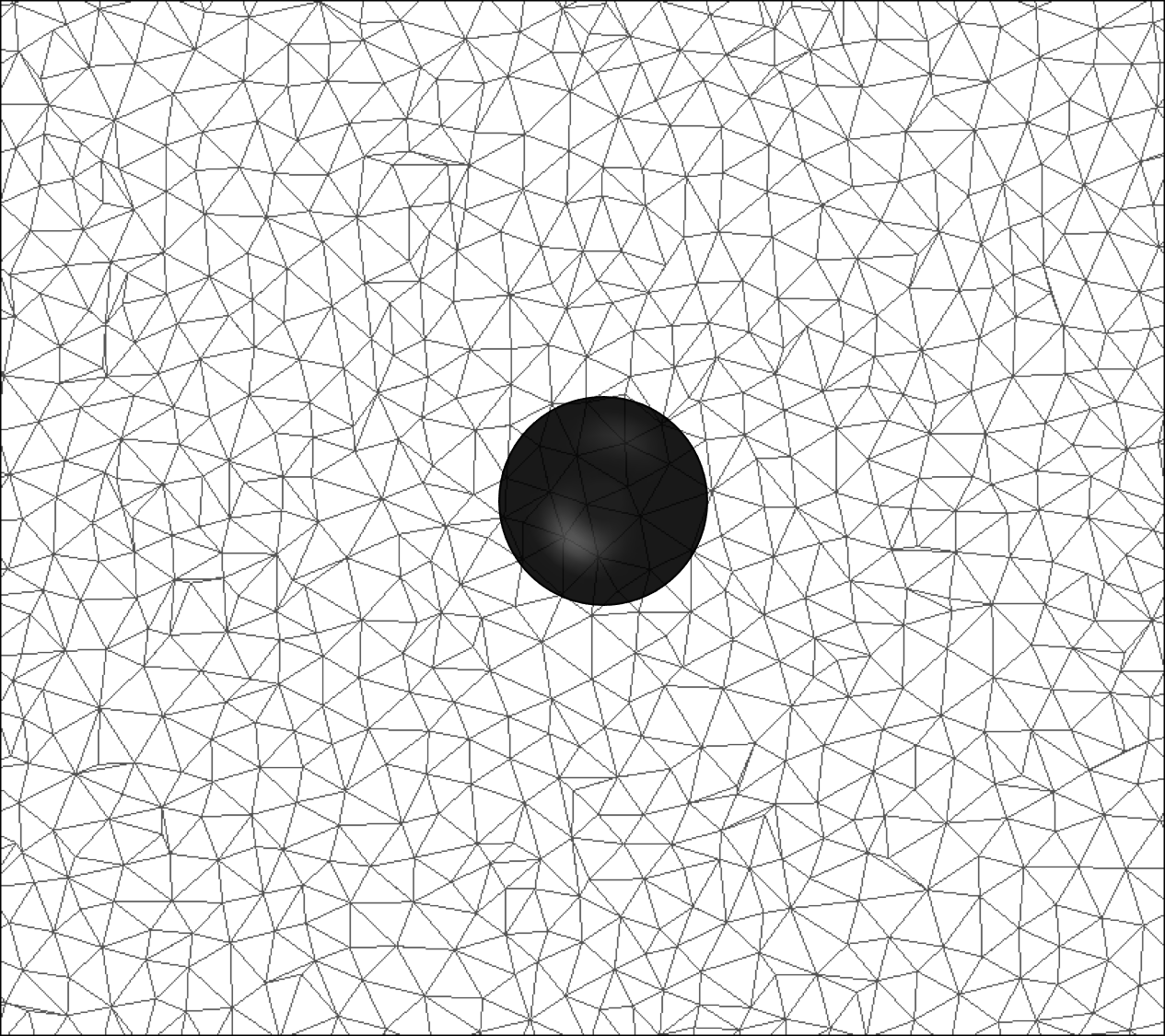}
     \includegraphics[trim={2.6inch 2.1inch 2.4inch 1.9inch},clip,scale=0.47]{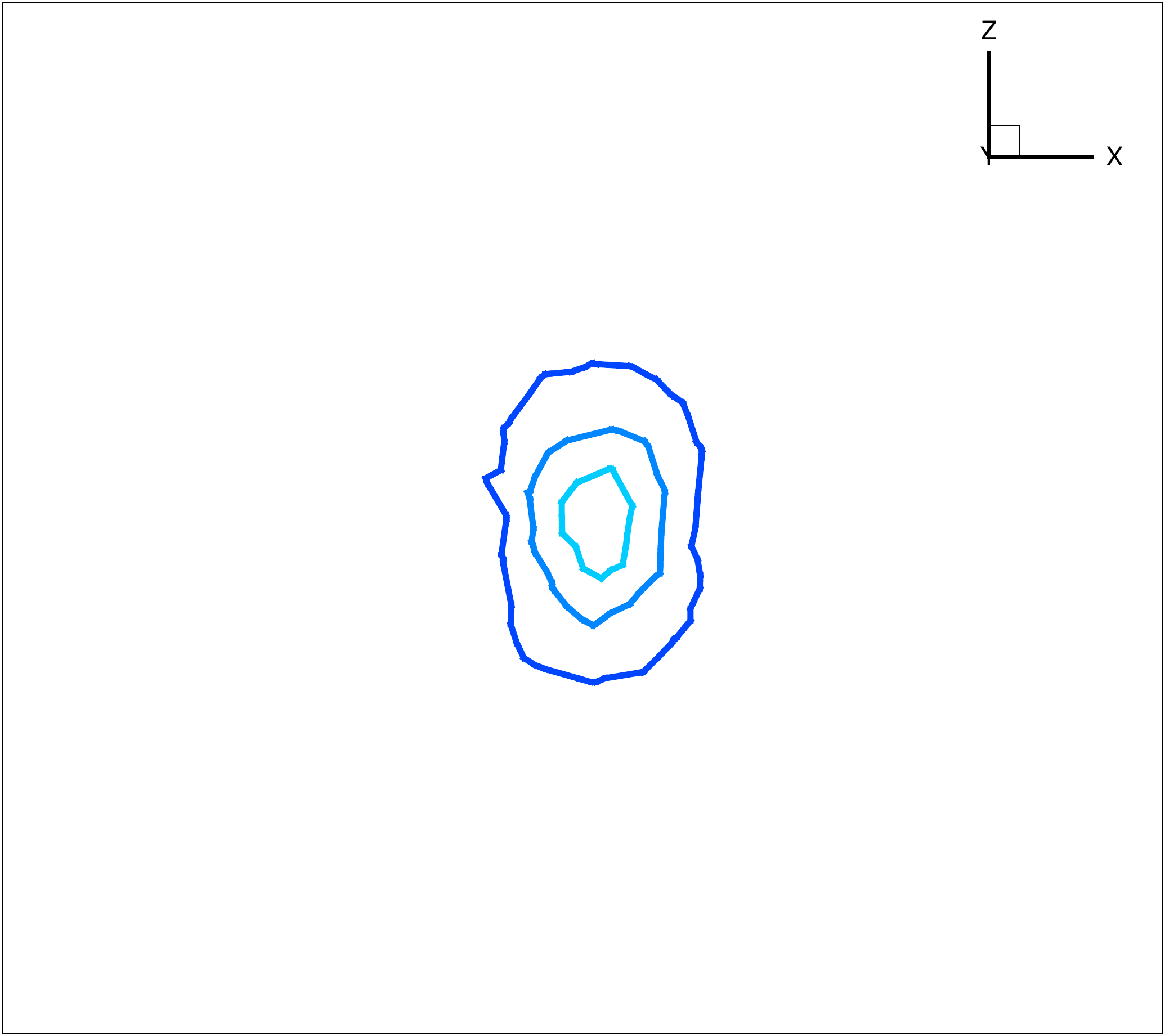}
    \includegraphics[trim={2.9inch 2.1inch 2.3inch 1.9inch},clip,scale=0.47]{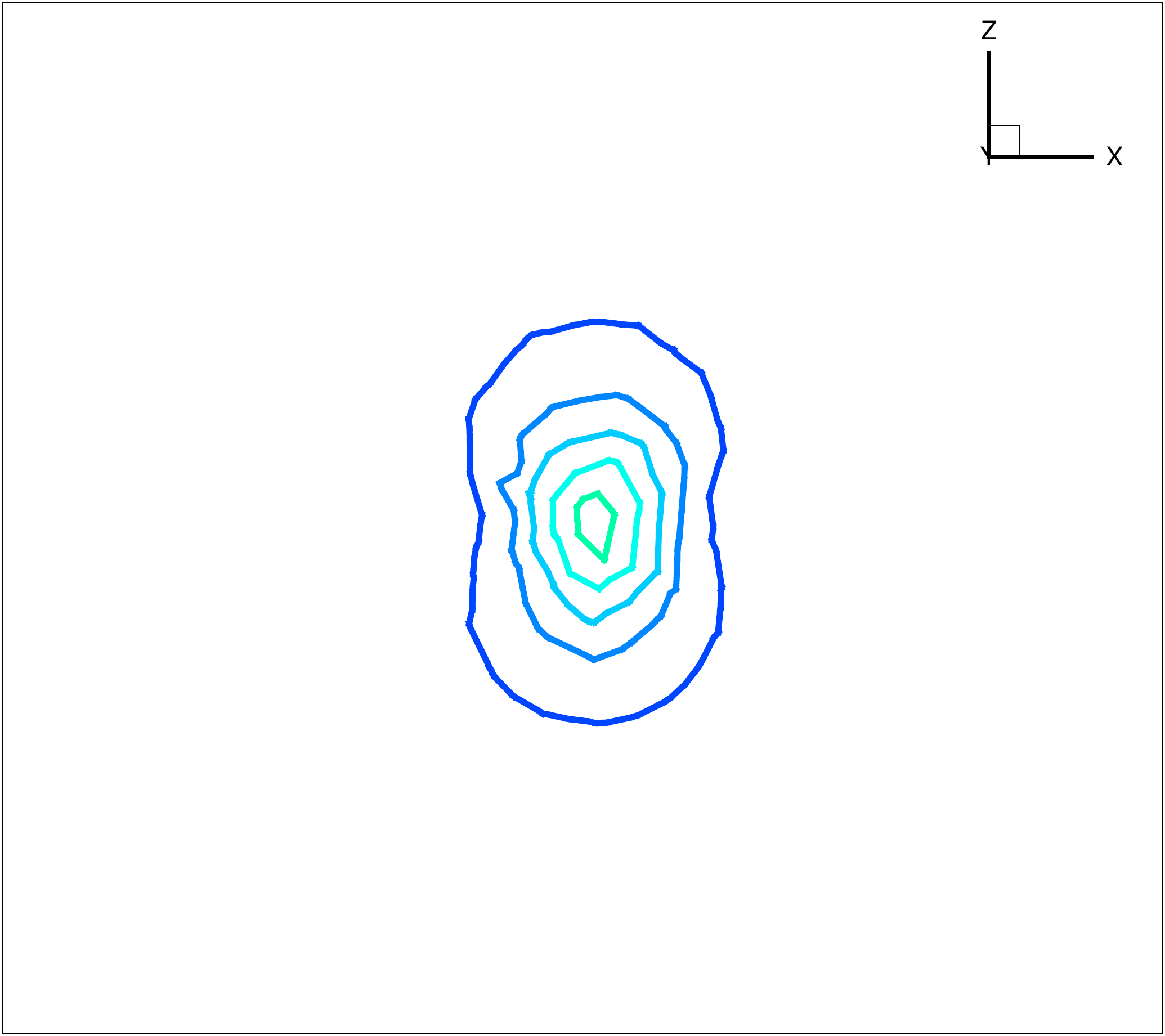}
    \includegraphics[trim={2.8inch 2.1inch 2.1inch 1.9inch},clip,scale=0.47]{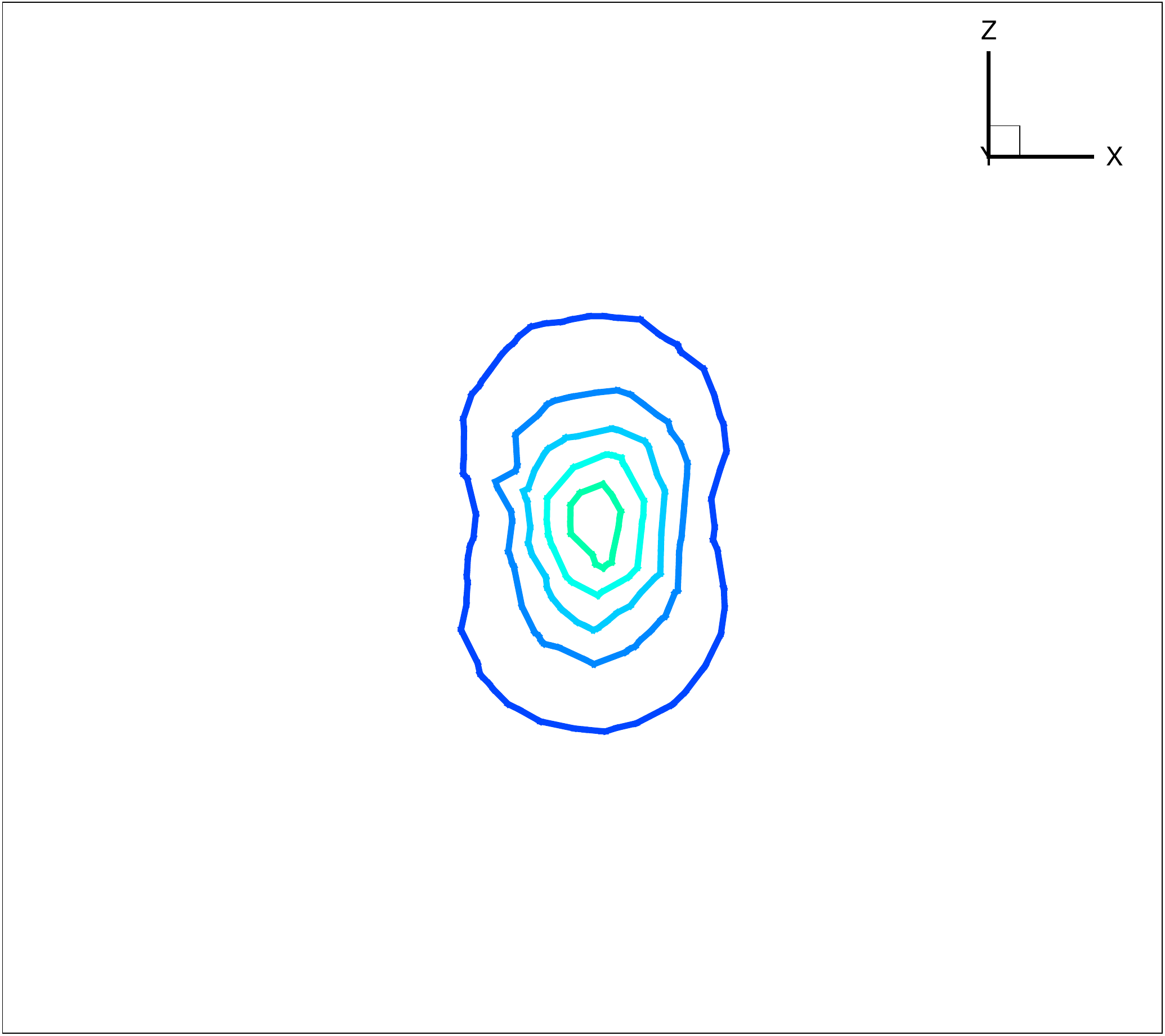}\\
    \caption{Contours of fluid velocity magnitude normalized by the particle settling speed at time instance of $2\tau_p$: particle and grid resolution (first column), uncorrected (second column), direct method correction (third column), and approximate method correction (fourth column). Results are based on case U2 (top row) and U9 (bottom row) from Tab.\ref{tab:unbounded_low_re} with $\Lambda=5.0$.} 
    \label{fig:countors}
\end{figure}

The performance of each method in capturing the effect of the particle on the fluid phase is illustrated in Fig.\ref{fig:countors} that pertains to cases U2 and U9 from Tab.\ref{tab:unbounded_low_re} with $\Lambda{=}5.0$. Contours of fluid velocity magnitude normalized by the particle settling velocity of each case are shown at the time instance of $2\tau_p$, from the initial release of the particle. The uncorrected scheme is compared against the corrected results using both direct as well as approximate methods. For the uncorrected scheme, the fluid phase experiences smaller velocity compared to the corrected results, owing to the smaller slip velocity that the particle samples in this scheme, resulting in a smaller drag force exerted to the flow. Concerning the predictions of the corrected schemes, both direct and approximate methods show nearly identical results in capturing the particle's effect on the background flow. The observations here imply stronger inter-phase coupling when the PP approach is corrected, which could potentially yield more accurate predictions for two-way coupled particle-laden flows. 

\begin{table}
\begin{center}
\def~{\hphantom{0}}
\begin{adjustbox}{width=\textwidth}
\begin{tabular} {lcccccc}
\hline
cell shape & case & $St$ & $\mathbf{\Lambda}$  &  \thead{uncorrected \\ $e^{\parallel}$ \quad \quad $e^{\perp}$ \quad \quad $e$} & \thead{corrected using \\ 
direct method\\ $e^{\parallel}$ \quad \quad $e^{\perp}$ \quad \quad $e$} & \thead{corrected using \\ 
approximate method\\ $e^{\parallel}$ \quad \quad $e^{\perp}$ \quad \quad $e$} \\ 
\hline
&&&&&&\\
\multirow{3}{*}{\includegraphics[scale=0.5]{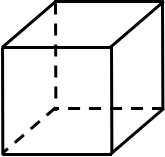}} & U1 & 10.0 & [1.0,1.0,1.0] & 54.2 \quad 0.077 \quad 54.2 & -0.056 \quad 0.0006 \quad 0.056 & 0.95 \quad 0.0689 \quad 0.95 \\ 
& U2 & 10.0 & [5.0,5.0,5.0] & 276.5 \quad 0.21 \quad 276.5 & -0.28 \quad 0.007 \quad 0.28 & 5.0 \quad 0.54 \quad 5.04\\
& U3 & 3.0 & [1.0,1.0,1.0] & 52.13 \quad 0.06 \quad 52.13 & -0.008 \quad 0.0003  \quad 0.008 & 0.08 \quad 0.1 \quad 0.17 \\
&&&&&&\\

\multirow{4}{*}{\includegraphics[scale=0.5]{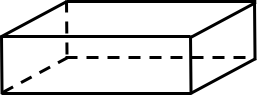}} & U4 & 10.0 & [5.0,0.5,0.5] & 35.42 \quad 2.86 \quad 35.54 & 0.0003 \quad 0.0003 \quad 0.0004 & 0.76 \quad 2.64 \quad 2.75 \\
& U5 & 10.0 & [4.0,2.0,0.2] & 30.26 \quad 4.56 \quad 30.61 & 0.0003 \quad 0.0005 \quad 0.0007 & 1.83 \quad 4.31 \quad 4.7\\
& U6 & 10.0 & [0.3,6.0,0.6] & 21.22 \quad 3.53 \quad 21.51 & 0.0004 \quad 0.0002 \quad 0.0004 & -2.31 \quad 3.42 \quad 4.13 \\
& U7 & 3.0 & [0.3,6.0,0.6] & 14.52 \quad 3.82 \quad 15.02 & 0.0006 \quad 0.0001 \quad  0.0006 & -2.28 \quad 3.85 \quad 4.48 \\
&&&&&&\\

\multirow{3}{*}{\includegraphics[scale=0.5]{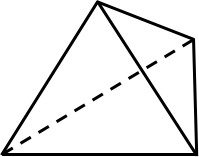}} & U8 & 10.0 & 1.0 & 25.55 \quad 0.44 \quad 25.56 & -0.06 \quad 0.06 \quad 0.11 & -3.01 \quad 0.48 \quad 3.06 \\
 & U9 & 10.0 & 5.0 & 129.66 \quad 2.17 \quad 129.68 & 0.65 \quad 0.39 \quad 0.81  & -17.75 \quad 2.77 \quad 18.03 \\
 & U10 & 3.0 & 1.0 & 25.39 \quad 0.60 \quad 25.40 & 0.10 \quad 0.11 \quad 0.16 & -3.63 \quad 0.62 \quad 3.70 \\
 &&&&&&\\
\hline
\end{tabular}
\end{adjustbox}
\caption{Listed are the percentage error in particle settling velocity $e^{||}$, drifting velocity $e^{\perp}$, and total velocity $e$, of a particle in an unbounded, quiescent flow. Results compare the uncorrected scheme to the direct and approximate correction methods for $Re_p{=}0.1$. Computational grids with different shapes and particle-to-gird size ratios are shown.}
\label{tab:unbounded_low_re}
\end{center}
\end{table}

\begin{figure}
    \centering
    \includegraphics[scale=0.57]{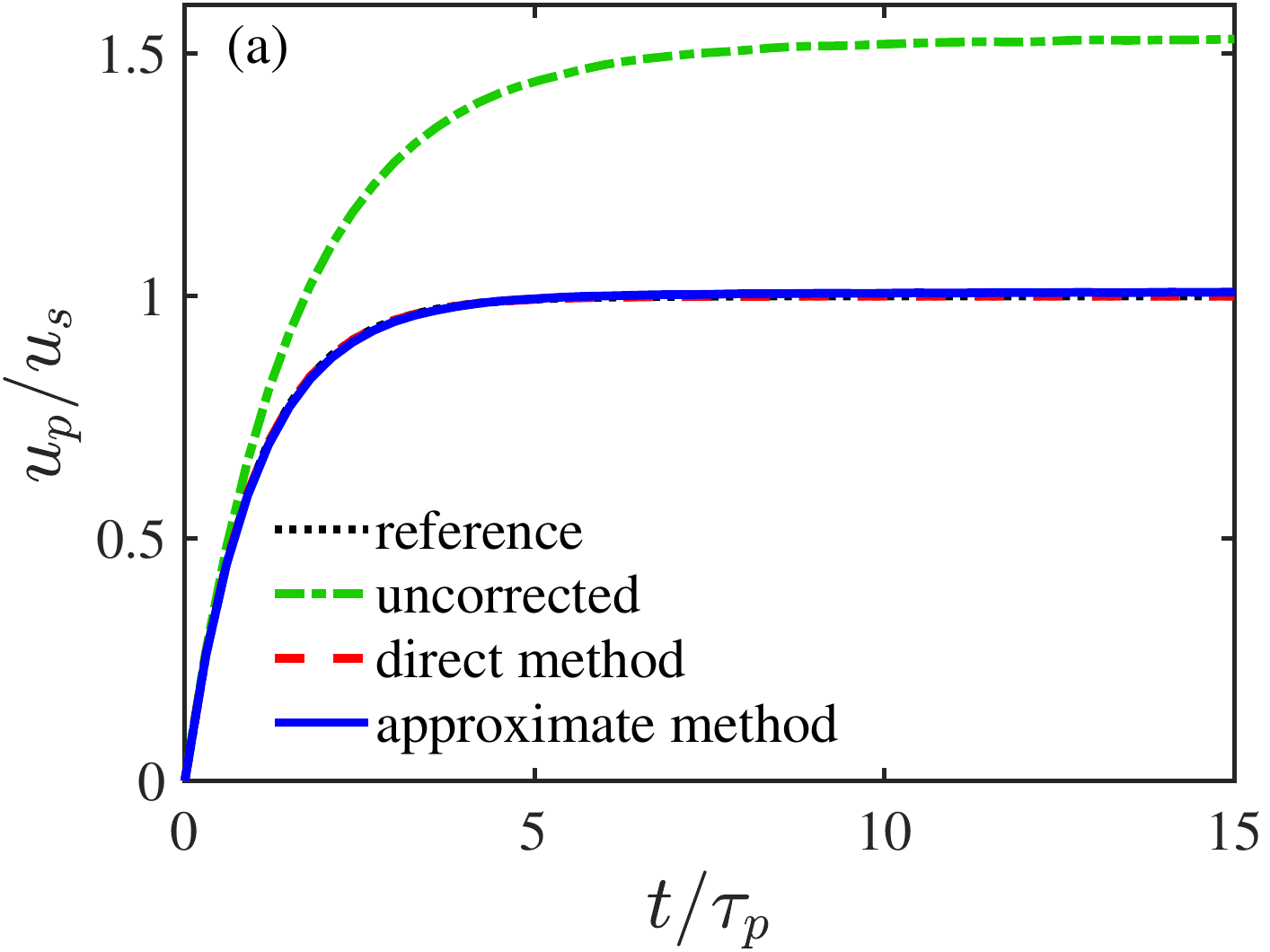}
    \includegraphics[scale=0.57]{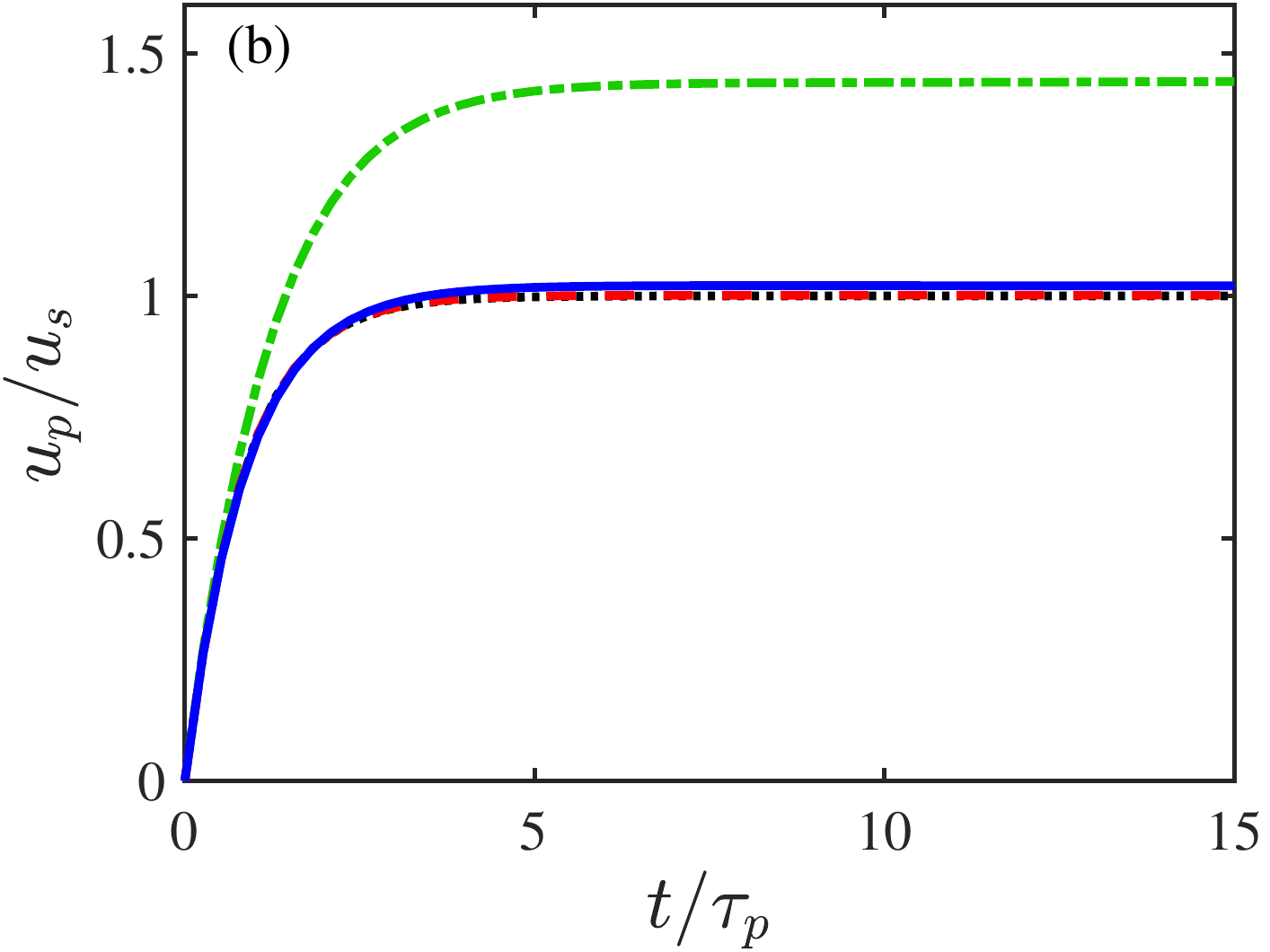}\vspace{0.06\textwidth}
    \includegraphics[scale=0.57]{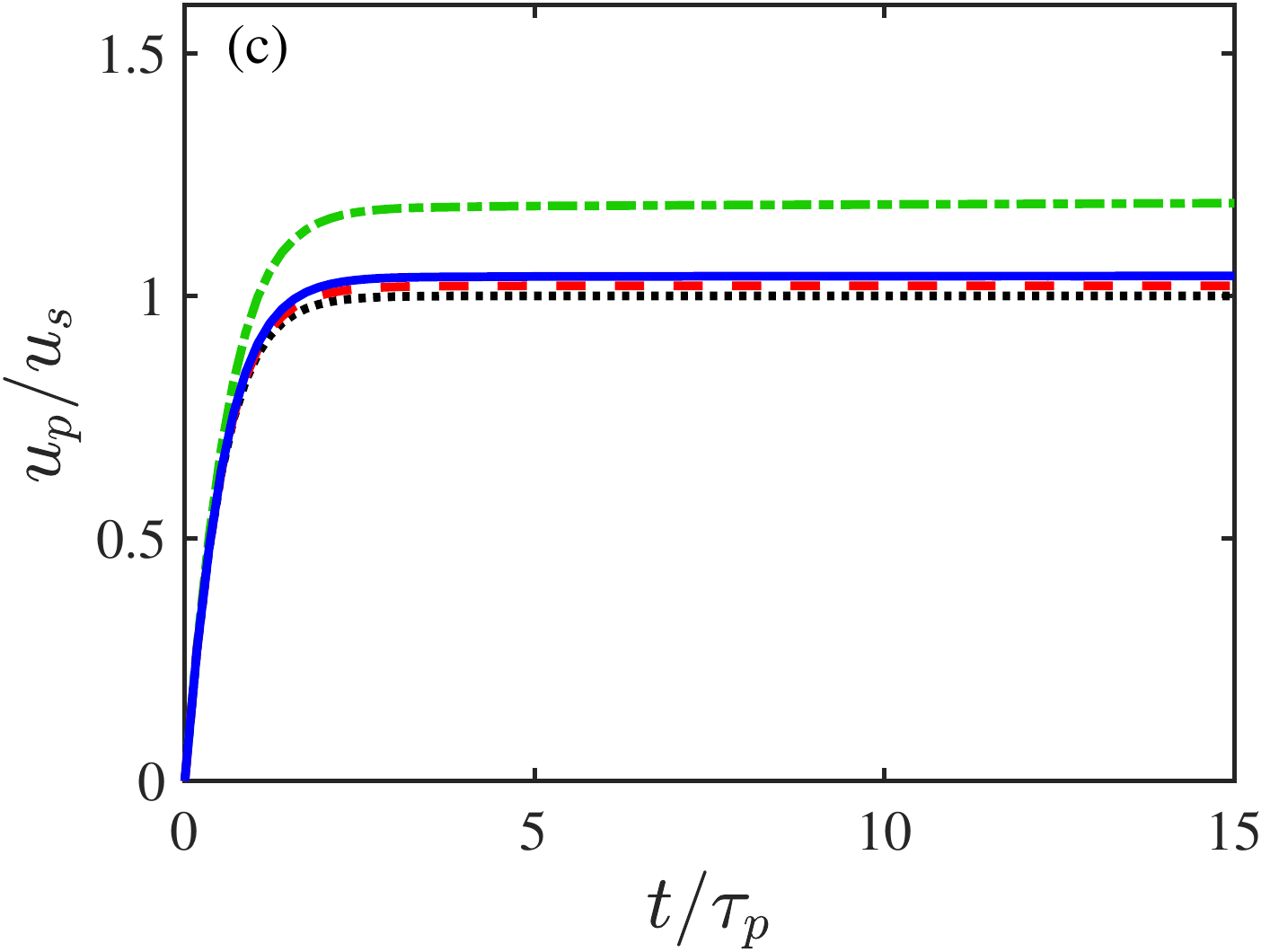}
    \includegraphics[scale=0.57]{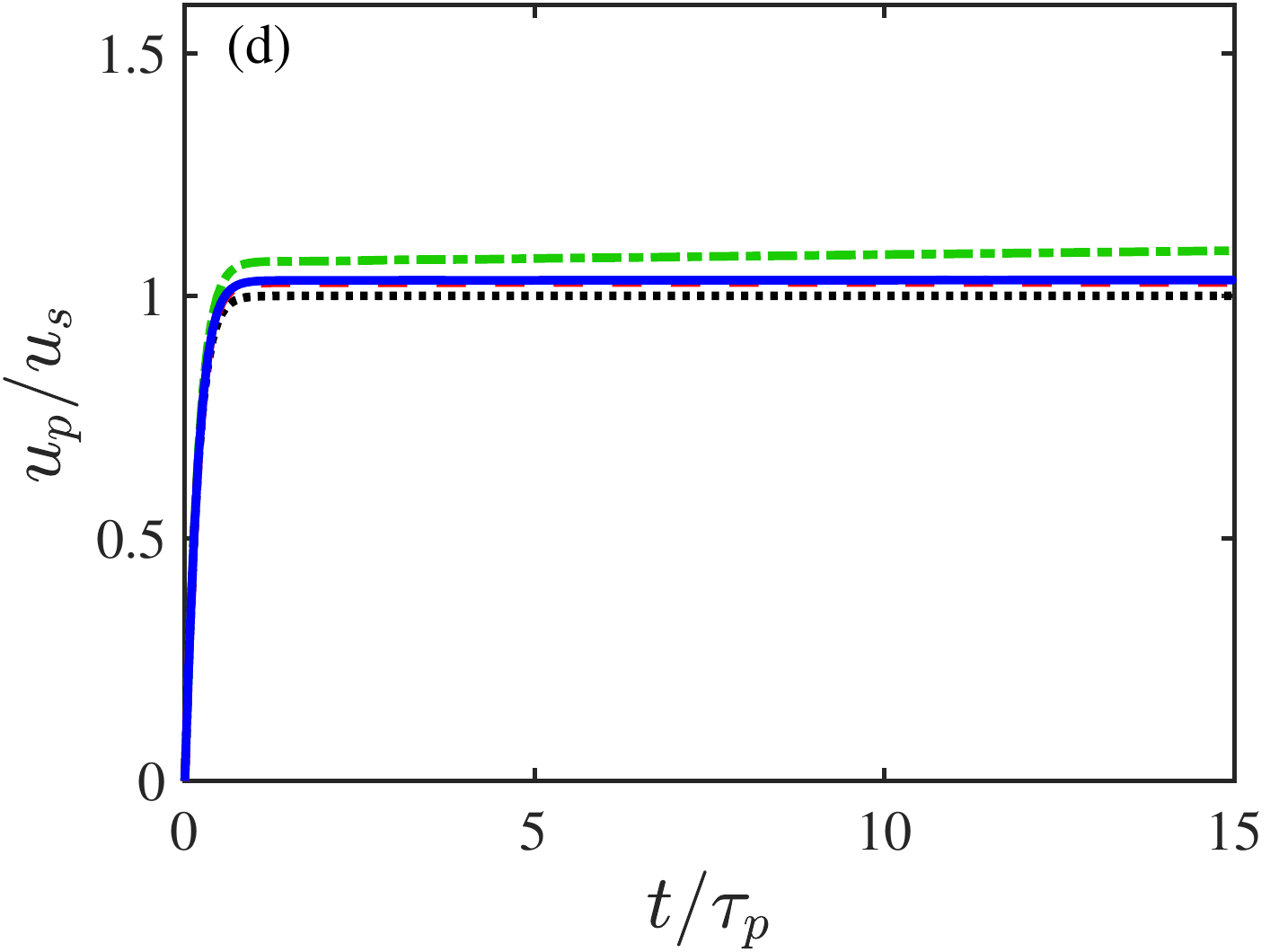}
    \caption{Temporal evolution of particle settling velocity in an unbounded, quiescent flow at (a): $Re_p{=}0.1$, (b): $Re_p{=}1.0$, (c): $Re_p{=}10.0$ and (d): $Re_p{=}100.0$. Results of the approximate method (solid blue), direct method (dashed red), uncorrected scheme (dash-dotted green) are compared against the reference velocity (dotted black). Results are based on cases U1 and UR1-UR3 from Tab. \ref{tab:unbounded_low_re} and \ref{tab:unbounded_high_re}, respectively. The settling velocity of $u_s{=}(1-\rho_f/\rho_p)\tau_p|\mathbf{g}|/f$ and partilce relaxation time of $\tau_p$ are used for normalizing the results.}
    \label{fig:unbounded_highRe}
\end{figure}

The predictive capability of the present models for a wide range of particle Reynolds number, typically encountered in various applications, is investigated next. Table \ref{tab:unbounded_high_re} provides settling in the range of $1{<}Re_p{<}100$ and using the same three grid configurations employed earlier. The first observation from the results here is that the errors in the uncorrected scheme decreases as particle Reynolds number increases, in line with the preceding works \citep{balachandar2019}, suggesting that the need for correction schemes becomes less important for $Re_p{>}100$. For instance, error of $44.67\%$ reduces to $11.16\%$ when $Re_p$ increases from $1$ to $100$ on the isotropic rectilinear grid. This is justified due to the fact that higher $Re_p$ particles move faster, and the residence time in their own disturbance field, created in the previous time step, becomes smaller than that of the slower particles, hence the lower disturbance. Moreover, \cite{balachandar2019} showed that as $Re_p$ increases, the region of maximum disturbance travels farther downstream so that the disturbed fluid velocity sampled at the particle location will be smaller for larger $Re_p$. 
Nevertheless, when the PP approach gets corrected by either methods, more accurate predictions are achieved for the studied range of $Re_p$. It should be noted that even though the correction may not be necessary for large $Re_p$ cases (e.g., see error of $1.35$ for uncorrected case of UR6), depending upon the particle-to-grid size ratio and grid anisotropy, the errors in uncorrected settling velocity could still be on the order of 10\% (see case UR3). For such cases, both methods are still effective in reducing the errors of the uncorrected scheme by an order of magnitude. Such a capability of the present models for reducing errors even for large $Re_p$ cases makes them more general to be employed without any restriction for a specific range of application. Figure \ref{fig:unbounded_highRe} illustrates the time dependent velocity of a single particle settling at different $Re_p$ predicted by the present correction schemes compared to the uncorrected PP approach and the reference. Settling velocity of $u_s{=}(1-\rho_f/\rho_p)\tau_p|\mathbf{g}|/f$ and particle time scale of $\tau_p$ are used for normalizing the results. 

\begin{table}
\begin{center}
\def~{\hphantom{0}}
\begin{adjustbox}{width=\textwidth}
\begin{tabular} {lcccccc}
\hline
cell shape & case & $Re_p$ & $\mathbf{\Lambda}$  &  \thead{uncorrected \\ $e^{\parallel}$ \quad \quad $e^{\perp}$ \quad \quad $e$} & \thead{Corrected using \\ 
Direct method\\ $e^{\parallel}$ \quad \quad $e^{\perp}$ \quad \quad $e$} & \thead{corrected using \\ 
approximate method\\ $e^{\parallel}$ \quad \quad $e^{\perp}$ \quad \quad $e$} \\ 
\hline
& & & & & \\
\multirow{3}{*}{\includegraphics[scale=0.5]{./figures/cube.png}} & UR1 & 1.0 & [1.0,1.0,1.0] & 44.67 \quad 0.12  \quad 44.67 & 0.16 \quad 0.007 \quad 0.16 & 2.18 \quad 0.03 \quad 2.18 \\
& UR2 & 10.0 & [1.0,1.0,1.0] & 19.9 \quad 0.58 \quad 19.9 & 2.08 \quad 0.03 \quad 2.08  & 4.26 \quad 0.66 \quad 4.31 \\
& UR3 & 100.0 & [1.0,1.0,1.0] & 11.15 \quad 0.21 \quad 11.16 & 2.82 \quad 0.03 \quad 2.82 & 3.44 \quad 0.24 \quad 3.45 \\
& & & & & \\

\multirow{3}{*}{\includegraphics[scale=0.5]{./figures/rectangle.png}} & UR4 & 1.0 & [0.3,6.0,0.6] & 20.32 \quad 3.79 \quad 20.67 & -0.003 \quad 0.001 \quad 0.003 & -1.77 \quad 3.92 \quad 4.31  \\
& UR5 & 10.0 & [0.3,6.0,0.6] & 7.53 \quad 1.47 \quad 7.67 & -0.02 \quad 0.01 \quad 0.03 & -0.27 \quad 1.62 \quad 1.65 \\
& UR6 & 100.0 & [0.3,6.0,0.6] & 1.34 \quad 0.22 \quad 1.35 & -0.56 \quad 0.02 \quad 0.56 & -0.64 \quad 0.22 \quad 0.68 \\
& & & & & \\

\multirow{3}{*}{\includegraphics[scale=0.5]{./figures/tetrahedral.png}} & UR7 & 1.0 & 1.0 & 15.32 \quad 0.47 \quad 15.33 & -0.07 \quad 0.03 \quad 0.1 & -3.07 \quad 0.42 \quad 3.10 \\
& UR8 & 10.0 & 1.0 & 7.17 \quad 0.15 \quad 7.17 & 0.48 \quad 0.03 \quad 0.48 & 0.52 \quad 0.17 \quad 0.56 \\
& UR9 & 100.0 & 1.0 & 2.33 \quad 0.03 \quad 2.33 & 0.49 \quad 0.01 \quad 0.49 & 0.59 \quad 0.03 \quad 0.59 \\
& & & & & & \\
\hline
\end{tabular}
\end{adjustbox}
\caption{Listed are the effect of $Re_p$ on the errors in settling, drifting and total velocity of a particle in an unbounded domain and predicted by direct, approximate and uncorrected schemes compared to the reference. Results are based on isotropic rectilinear grid, anisotropic rectilinear grid as well as a tetrahedral unstructured grid. $St=10.0$ is shared among the cases.}
\label{tab:unbounded_high_re}
\end{center}
\end{table}

\subsection{Settling parallel to a wall}
In this part, the capability of the present models for wall-bounded regimes is evaluated. For test cases here, an additional non-dimensional parameter that is the normalized wall distance from the bottom of the particle is defined as, 

\begin{equation}
    \delta_p = \frac{x_{2,p}}{d_p}-0.5,
    \label{eq:delta_p}
\end{equation}


\noindent wherein $x_{2,p}$ is the wall-normal distance from the center of the particle (wall is assumed in $x_{1}$--$x_{3}$ plane). Concerning complex geometries, computing this distance to the nearest wall might not be straightforward and would have to be investigated in the future. 

Settling velocity of a particle at various wall distances is carried out using the present methods and on the three aforementioned computational grids. For the studied cases, a particle that is initially stationary and located at a given $\delta_p$, released to reach its settling velocity under a gravity vector of $\mathbf{g}{=}[\exp(1),0,(1+\sqrt{5})/2]/|\mathbf{g}|$ that guarantees the particle's motion on a plane parallel to the wall. In reality, the particle experiences a lateral force \citep{vasseur1977,takemura2003} that is neglected in this study to isolate the parallel motion. The particle's equation of motion in the presence of wall still follows Eq. \ref{eq:dup_dt} with the adjustment factor of $f$ that accounts for the wall effects on the particle's drag coefficient. Concerning this factor, the empirical expression derived by \cite{zeng2009} is employed that covers a wide range of $\delta_p$ and $Re_p$ as, 

\begin{equation}
    f^{||}(\delta_p,Re_p) = f^{||}_1(\delta_p)f^{||}_2(\delta_p,Re_p),
    \label{eq:f_parallel}
\end{equation}

\noindent where, 

\begin{equation}
    f^{||}_1(\delta_p) = \left[1.028 - \frac{0.07}{1+4\delta^2_p} - \frac{8}{15}\log\left(\frac{270\delta_p}{135+256\delta_p}\right)\right];
\end{equation}

\begin{equation}
    f^{||}_2(\delta_p,Re_p) = \left[ 1+0.15\left(1-\exp\left(-\sqrt{\delta_p}\right)\right)Re_p^{\left(0.687+0.313\exp\left(-2\sqrt{\delta_p}\right)\right)}\right].
\end{equation}

\noindent $f^{||}_1(\delta_p)$ captures wall effects on the Stokes drag for zero $Re_p$, that approaches unity when $\delta_p{\rightarrow}\infty$. The second term, $f^{||}_2(\delta_p,Re_p)$, however, handles the wall-modified finite Reynolds number effect on the Stokes drag coefficient that converts to the Schiller-Naumman adjustment factor (Eq. \ref{eq:schiller-Naumman}) when particle travels sufficiently away from the wall. 

Using the correction factor expressed above, the particle's equation of motion (Eq.\ref{eq:dup_dt}) is solved using the corrected and uncorrected PP approaches. Following the metrics presented in the preceding subsection, the errors in settling, drifting and total velocity of the particle are measured in comparison to those of the one-way coupled simulations that serves as the reference. Table \ref{tab:parallel_low_re} shows these errors for settling velocity of a particle with $Re_p{=}0.1$ and $St{=}10$ on (i) isotropic rectilinear grid (set A), (ii) anisotropic rectilinear grid (set B), and (iii) tetrahedral unstructured grid (set C). Each grid has six cases corresponding to settling at different $\delta_p$, that covers a wide range of distances from the wall.

The first observation from Tab. \ref{tab:parallel_low_re} is that the uncorrected scheme produces significantly large errors in predicting particle velocity at all wall distances, with slightly larger values near the wall, consistent with the observations of \cite{pakseresht2020}. It is imperative to mention that the reported errors here in this work are slightly smaller than those of \cite{pakseresht2020}, potentially due to different Euler-Lagrange interpolation schemes employed in this study as compared to their tri-linear interpolation. When the direct and approximate correction methods used to obtain undisturbed fluid velocity, the errors reduce by one or two orders of magnitude compared to the uncorrected scheme. Of specific interest is the results obtained from the approximate method that shows same order of accuracy as those reported by \cite{pakseresht2020} for rectilinear grids. Different grid configurations used in the present work, including unstructured grids, show the applicability of the present methods for more complex geometries that are encountered in the real world applications. 

\begin{table}
\begin{center}
\def~{\hphantom{0}}
\begin{adjustbox}{width=\textwidth}
\begin{tabular} {lcccccc}
\hline
cell shape & case & $\delta_p$& $\mathbf{\Lambda}$ &  \thead{uncorrected \\ $e^{\parallel}$ \quad \quad $e^{\perp}$ \quad \quad $e$} & \thead{corrected using \\ direct method\\ $e^{\parallel}$ \quad \quad $e^{\perp}$ \quad \quad $e$} & \thead{corrected using\\ approximate method\\ $e^{\parallel}$ \quad \quad $e^{\perp}$ \quad \quad $e$} \\ 
\hline
&&&&&& \\
\multirow{5}{*}{\includegraphics[scale=0.5]{./figures/cube.png}} & A1 & 0.05 & [1.0,1.0,1.0] & 71.32 \quad 0.03 \quad 71.32 & -1.38 \quad 0.005 \quad 1.38 & 3.99 \quad 0.09 \quad 3.99 \\ 
& A2 & 0.5 & [1.0,1.0,1.0] & 42.53 \quad 0.02 \quad 42.53 & -0.21 \quad 0.0008 \quad 0.21 & 0.89 \quad 0.04 \quad 0.9 \\
& A3 & 1.0 & [1.0,1.0,1.0] & 45.81 \quad 0.02 \quad 45.81 & -0.05 \quad 0.0005 \quad 0.05 & 0.27 \quad 0.04 \quad 0.27 \\
& A4 & 2.0 & [1.0,1.0,1.0] & 49.73 \quad 0.02 \quad 49.73 & -0.13 \quad 0.0006 \quad 0.13 & -0.02 \quad 0.04 \quad 0.05 \\
& A5 & $\infty$ & [1.0,1.0,1.0] & 54.44 \quad 0.04 \quad 54.44 & 0.03 \quad 0.0008 \quad 0.06 & 1.14 \quad 0.05 \quad 1.15 \\
&&&&&\\

\multirow{5}{*}{\includegraphics[scale=0.5]{./figures/rectangle.png}} & B1 & 0.05 & [0.3,6.0,0.6] & 29.28 \quad 1.62 \quad 29.33 & 0.08 \quad 0.00 \quad 0.08 & 7.6 \quad 1.62 \quad 7.77 \\
& B2 & 0.5 & [0.3,6.0,0.6] & 24.18 \quad 1.71 \quad 24.24 & 0.14 \quad 0.00 \quad 0.14 & 5.64 \quad 1.72 \quad 5.9 \\
& B3 & 1.0 & [0.3,6.0,0.6] & 26.73 \quad 2.26 \quad 26.82 & 0.27 \quad 0.00 \quad 0.27 & 5.32 \quad 2.28 \quad 5.79 \\
& B4 & 2.0 & [0.3,6.0,0.6] & 28.8 \quad 2.61 \quad 28.91 & 0.53 \quad 0.001 \quad 0.53 & 4.44 \quad 2.64 \quad 5.17 \\
& B5 & $\infty$ & [0.3,6.0,0.6] & 28.53 \quad 2.37 \quad 28.63 & 1.25 \quad 0.02 \quad 1.25 & 3.59 \quad 2.46 \quad 4.36 \\
&&&&&\\

\multirow{5}{*}{\includegraphics[scale=0.5]{./figures/tetrahedral.png}} & C1 & 0.05 & 1.0 & 30.28 \quad 0.50 \quad 30.28 &  0.53 \quad 0.29 \quad 0.64 & -1.47 \quad 0.59 \quad 1.68 \\
& C2 & 0.5 & 1.0 & 19.13 \quad 0.34 \quad 19.14 & 0.18 \quad 0.14 \quad 0.25 & -1.21 \quad 0.33  \quad 1.28 \\
& C3 & 1.0 & 1.0 & 18.60 \quad 0.29 \quad 18.60 & -0.06 \quad 0.08 \quad 0.13 & -2.17 \quad 0.29 \quad 2.21 \\
& C4 & 2.0 & 1.0 & 21.63 \quad 0.38 \quad 21.63 & -0.32 \quad 0.06 \quad 0.33 & -2.75 \quad 0.45 \quad 2.8 \\
& C5 & $\infty$ & 1.0 & 25.99 \quad 0.16 \quad 25.99  & -0.19 \quad 0.04 \quad 0.19 & -3.16 \quad 0.2 \quad 3.17 \\
&&&&&& \\
\hline
\end{tabular}
\end{adjustbox}
\caption{Percentage errors in the settling velocity, $e^{||}$, drifting velocity, $e^{\perp}$, and total velocity, $e$, of a single particle in parallel motion to a no-slip wall on different grids for $St{=}10$ and $Re_p{=}0.1$ and for a range of $\delta_p$.}
\label{tab:parallel_low_re}
\end{center}
\end{table}

\begin{table}
\begin{center}
\def~{\hphantom{0}}
\begin{adjustbox}{width=\textwidth}
\begin{tabular} {lcccccc}
\hline
cell shape & case & $Re_p$ & $\mathbf{\Lambda}$ &  \thead{uncorrected \\ $e^{\parallel}$ \quad \quad $e^{\perp}$ \quad \quad $e$} & \thead{corrected using \\ direct method\\ $e^{\parallel}$ \quad \quad $e^{\perp}$ \quad \quad $e$} & \thead{corrected using\\ approximate method\\ $e^{\parallel}$ \quad \quad $e^{\perp}$ \quad \quad $e$} \\ 
\hline
&&&&&& \\
\multirow{3}{*}{\includegraphics[scale=0.5]{./figures/cube.png}} & WR1 & 1.0 & [1.0,1.0,1.0] & 65.21 \quad 0.11 \quad 65.21 & -0.92 \quad 0.005 \quad 0.92 & 3.86 \quad 0.01 \quad 3.86 \\
& WR2 & 10.0 & [1.0,1.0,1.0] & 31.92 \quad 0.86 \quad 31.93 & 4.29 \quad 0.09 \quad 4.29 & 8.06 \quad 0.91 \quad 8.11 \\
& WR3 & 100.0 & [1.0,1.0,1.0] & 16.48 \quad 0.37 \quad 16.48 & 10.72 \quad 0.24 \quad 10.72 & 11.08 \quad 0.4 \quad 11.09 \\
&&&&&& \\

\multirow{3}{*}{\includegraphics[scale=0.5]{./figures/rectangle.png}} & WR4 & 1.0 & [0.3,6.0,0.6] & 28.21 \quad 1.58 \quad 28.25 & 0.04 \quad 0.001 \quad 0.04 & 7.01 \quad 1.60 \quad 7.19 \\
& WR5 & 10.0 & [0.3,6.0,0.6] & 19.28 \quad 0.84 \quad 19.30 & 0.11 \quad 0.004 \quad 0.11 & 3.86 \quad 0.95 \quad 3.98 \\
& WR6 & 100.0 & [0.3,6.0,0.6] & 9.98 \quad 0.1 \quad 9.98 & 0.03 \quad 0.007 \quad 0.03 & 1.23 \quad 0.13 \quad 1.23 \\
&&&&&& \\
\multirow{3}{*}{\includegraphics[scale=0.5]{./figures/tetrahedral.png}} & WR7 & 1.0 & 1.0 & 25.94 \quad 0.6 \quad 25.95 & 0.43 \quad  0.38 \quad 0.59 & -2.68 \quad 0.7 \quad 2.8 \\
& WR8 & 10.0 & 1.0 & 12.09 \quad 0.18 \quad 12.09 & 0.21 \quad 0.05 \quad 0.22 & -0.4 \quad 0.22 \quad 0.51 \\
& WR9 & 100.0 & 1.0 & 3.72 \quad 0.04 \quad 3.72 & -0.1 \quad 0.003 \quad 0.1 & -0.04 \quad 0.03 \quad 0.06 \\
&&&&&& \\
\hline
\end{tabular}
\end{adjustbox}
\caption{Effect of $Re_p$ on the particle settling velocity at $\delta_p{=}0.05$ predicted by the present models in comparison with uncorrected scheme showing errors in settling, $e^{||}$, drifting, $e^{\perp}$ and total, $e$, velocities.}
\label{tab:parallel_highRe}
\end{center}
\end{table}

In order to test the present methods for higher $Re_p$ in wall-bounded regimes, settling velocity of a particle near a no-slip wall is computed for $1.0{<}Re_p{<}100$. Table~\ref{tab:parallel_highRe} gives the results for particle settling at $\delta_p{=}0.05$, for which the errors in the settling velocity were observed to be more remarkable in the preceding part. Consistent with the unbounded cases (see Tab. \ref{tab:unbounded_high_re}), as $Re_p$ increases the error in the particle velocity decreases and the need for correcting the PP approach diminishes. As an example, for the particle settling on the unstructured grid, the total error of $25.95$ for $Re_p{=}1.0$ (case WR7) decreases to $3.72$ when $Re_p$ increases to $100$ (case WR9). Nevertheless, both direct and approximate models are able to reduce the errors even for large $Re_p$ by an order of magnitude. Not shown here, similar results were obtained for settling at other wall distances with large $Re_p$.

\subsection{Free falling normal to a wall}
This section tests the ability of the present methods for recovering the undisturbed fluid velocity for particles in wall-normal motion. Freely falling particle toward a no-slip wall is studied. In this configuration, the drag coefficient of the particle increases as it approaches the wall, owing to the wall lubrication effect. Accordingly, for the wall adjustment factor to the particle's drag coefficient of this part, the asymptotic expressions derived by \cite{brenner1961,cox1967} as,

\begin{equation}
f^{\perp}(\delta_p) = 
\begin{cases}
 1 +\left( \frac{0.562}{1+2\delta_p}\right) \quad \text{for} \quad \delta_p>1.38 \quad \text{\citep{brenner1961}}\\
 \frac{1}{2\delta_p} \left( 1+ 0.4\delta_p\log\left(\frac{1}{2\delta_p}\right) + 1.94\delta_p \right) \quad \text{for} \quad \delta_p<1.38 \quad \text{\citep{cox1967},}\\
\end{cases}
\label{eq:brener_asymptotic}
\end{equation}

\noindent are employed that include two parts depending on the wall normal distance of the particle. This adjustment factor is used in Eq. \ref{eq:dup_dt} to compute the particle's equation of motion. 

Following the work of \cite{pakseresht2020}, a particle that is initially stationary and located at an arbitrary $\delta_p{=}7$, falls under gravity vector of $\mathbf{g}{=}[0,-1,0]$, and its velocity and wall-normal distance are measured as a function of time. Table~\ref{tab:error_normal} lists cases performed on various grid configurations, different particle Reynolds numbers in the range of $0.1{<}Re^{Stk}_p{<}100$ and two particle Stokes number of $St=3.0$ and $St=10.0$. For the computations of this part, the particle Reynolds number is defined based on the unbounded Stokes regime, $Re^{Stk}_p$, expressed by Eq. \ref{eq:Rep_Stk}. It should be noted that the drag expression provided by Eq. \ref{eq:brener_asymptotic} is only valid for $Re_p{<}0.1$, however, it is still employed for larger $Re_p$ cases just for numerical demonstration without advocating its use for $Re_p{>}0.1$.
The error for each method is measured based on the total time that the particle requires to reach $\delta_p{=}0.5$ in comparison to the reference value, $t_{ref}$, that is obtained based on the one-way coupled simulation. The deviation of each scheme from the reference is quantified based on the relative error as,

\begin{equation}
    e_n = \frac{t- t_{ref}}{t_{ref}}.
    \label{error_free_fall}
\end{equation}

Table~\ref{tab:error_normal} shows that for the studied cases, the uncorrected scheme yields negative errors revealing the fact that the particle in this scheme experiences smaller drag force, accelerates faster and reaches the wall distance of interest much quicker than it would in reality. When the PP approach is corrected using the direct method, however, small errors of ${\mathcal O}(0.1)$ are achieved that shows the successful predictions of this method for the studied range of flow parameters and the grid configurations. For the approximate method, the errors are reduced to smaller values as well, however, for the highly skewed anisotropic rectilinear grid, this method results in errors on the same order of magnitude of the uncorrected scheme. This is attributed to (i) the response of a fluid to a source in a control volume, may differ based on the shape of the control volume owing to the numerical approach used in solving the governing equation, (ii) even with anisotropic grids, the distribution of the particle reaction force is done to the nearest neighbors of the control volume which could be asymmetric with high aspect ratio grids, and (iii) for particle motion normal to a no-slip wall, the pressure distribution on the particle surface is asymmetric, and thus a simple approach to model the pressure gradient in the approximate method potentially needs to be modified. 
Concerning the effect of $Re_p$, it is observed that for the studied computational grids, as $Re_p$ increases the error in the uncorrected scheme decreases, consistent with the previous observations. 

\begin{table}
\begin{center}
\def~{\hphantom{0}}
\begin{adjustbox}{width=1.0\textwidth}
\begin{tabular} {lccccccc}
\hline
cell shape & case & $Re^{Stk}_p$ & $St$ & $\mathbf{\Lambda}$ &  \thead{uncorrected \\ $e_n$} & \thead{corrected using \\ direct method \\ $e_n$} & \thead{corrected using \\ approximate method \\ $e_n$} \\ 
\hline
&&&&&&&\\
\multirow{5}{*}{\includegraphics[scale=0.5]{./figures/cube.png}} & N1 & 0.1 & 3.0 & [1.0,1.0,1.0] & -28.12 & 0.40 & 9.28 \\
& N2 & 0.1 & 10 & [1.0,1.0,1.0] & -22.71 & 0.12 & 6.52 \\
& N3 & 1.0 & 10 & [1.0,1.0,1.0] & -6.56 & -0.28 & 0.56 \\
& N4 & 10.0 & 10 &  [1.0,1.0,1.0] & -1.29 & -0.07 & 0.10 \\
& N5 & 100.0 & 10 &  [1.0,1.0,1.0] & -0.2 & 0.00 & 0.04 \\
&&&&&&&\\

\multirow{5}{*}{\includegraphics[scale=0.5]{./figures/rectangle.png}} & N6 & 0.1 & 3.0 & [0.3,6.0,0.6] & -8.66 & -0.96 & 20.08 \\
& N7 & 0.1 & 10 &  [0.3,6.0,0.6] & -8.65 & -0.96 & 20.05 \\
& N8 & 1.0 & 10 & [0.3,6.0,0.6] & -5.92 & -0.26 & 14.28 \\
& N9 & 10.0 & 10 & [0.3,6.0,0.6] & -0.74 & 0.00 & 1.36 \\
& N10 & 100.0 & 10 & [0.3,6.0,0.6] & -0.04 & 0.00 & 0.09 \\
&&&&&&&\\

\multirow{5}{*}{\includegraphics[scale=0.5]{./figures/tetrahedral.png}} & N11 & 0.1 & 3.0 & 1.0 & -13.99 & 0.53 & 9.82 \\
& N12 & 0.1 & 10 & 1.0 & -12.04 & 0.37 & 7.90 \\
& N13 & 1.0 & 10 & 1.0 & -3.09 & 0.07 & 1.33 \\
& N14 & 10.0 & 10 & 1.0 & -0.54 & -0.02 & 0.15 \\
& N15 & 100.0 & 10 & 1.0 & -0.07 & 0.00 & 0.02  \\
&&&&&&&\\

\hline
\end{tabular}
\end{adjustbox}
\caption{Percentage errors in the time that a particle initially located at $\delta_p{=}7.0$ requires to reach $\delta_p{=}0.5$ for various grid configurations and different $Re^{Stk}_p$ and $St$ with and without correction schemes.}
\label{tab:error_normal}
\end{center}
\end{table}

\begin{figure}
    \centering
    \includegraphics[scale=0.56]{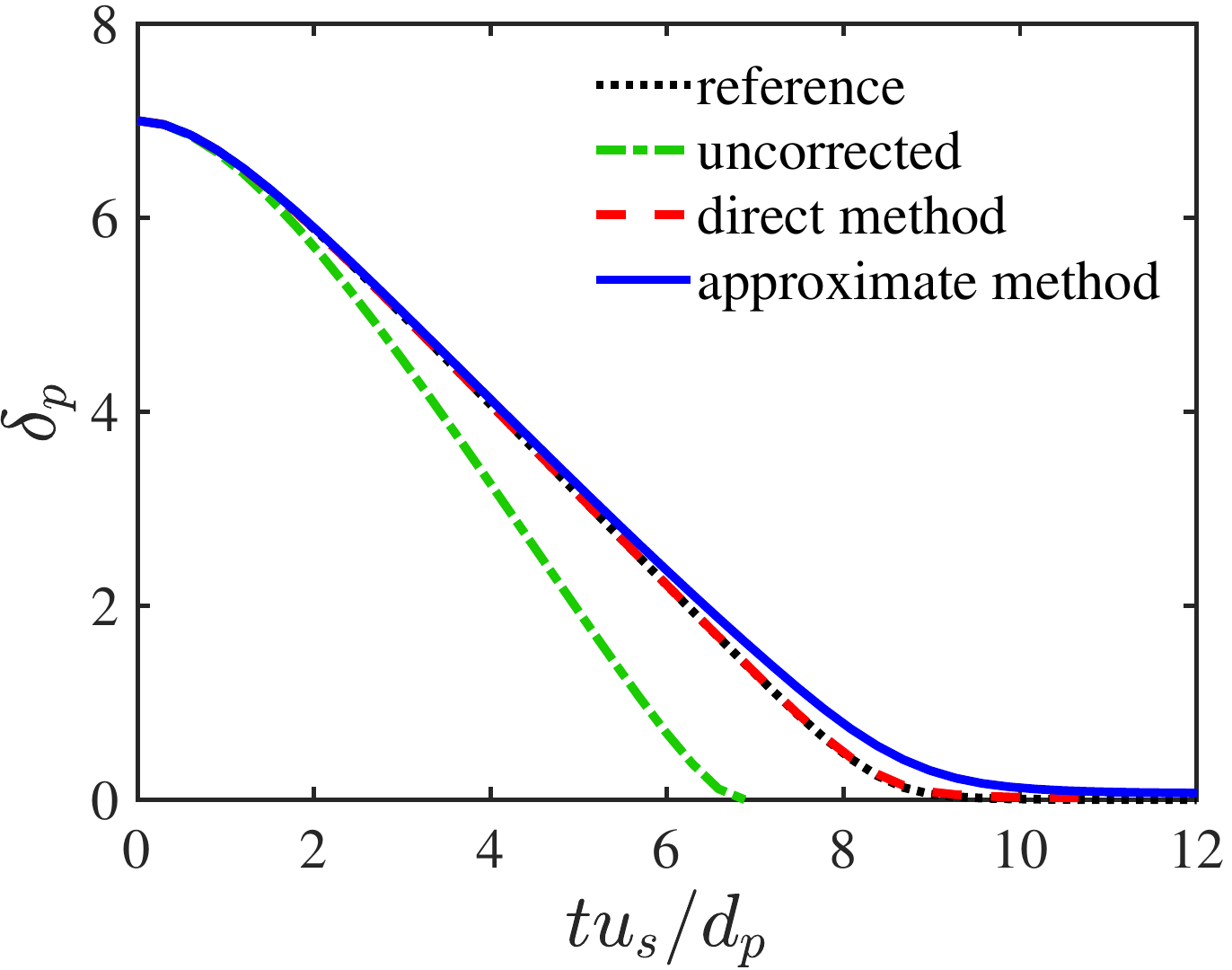}\hspace{0.01\textwidth}
    \includegraphics[scale=0.56]{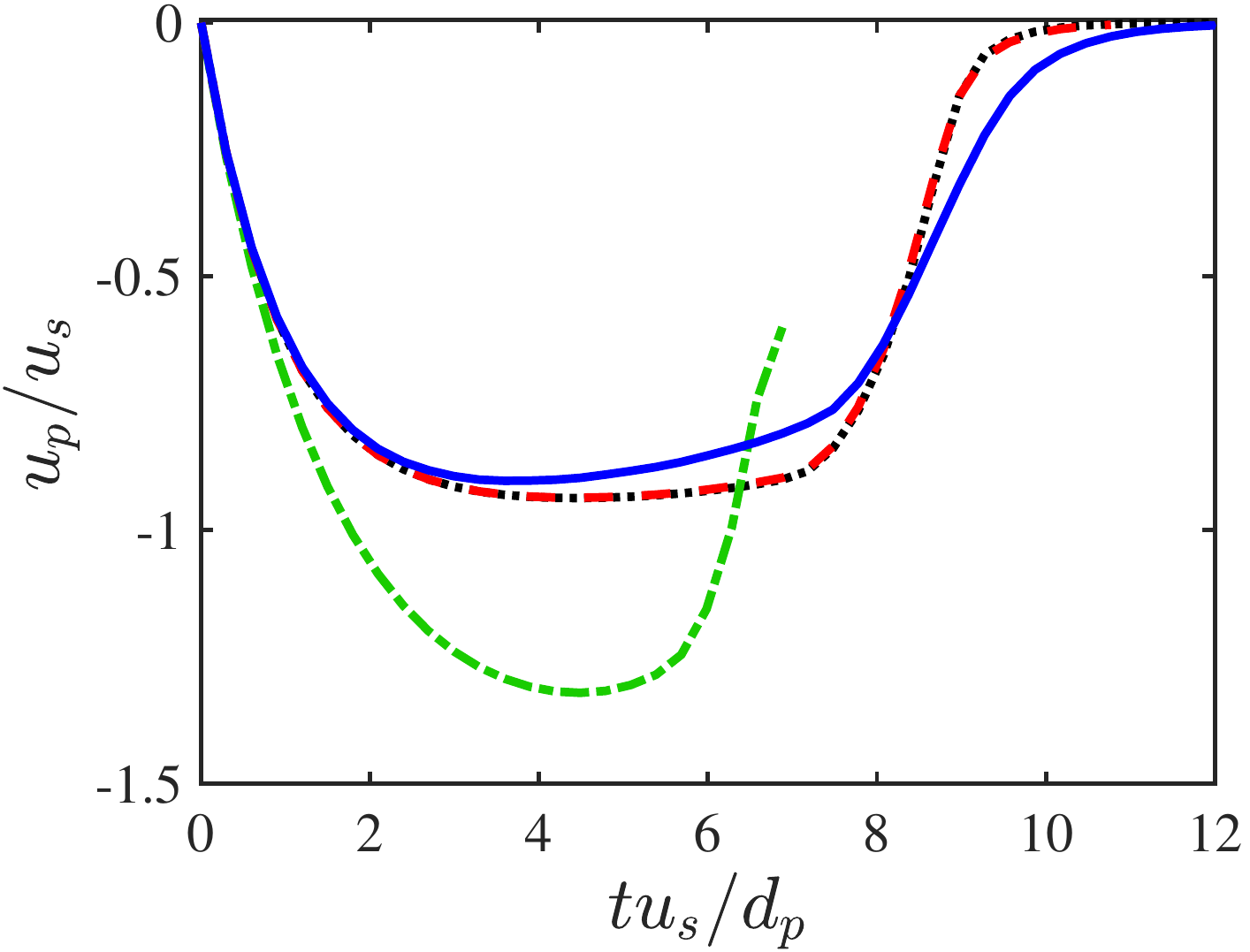}
    \caption{Normalized wall distances (left) and velocity (right) of a particle in moving normal to a no-slip wall. Results pertain to case N2 from Tab. \ref{tab:error_normal} comparing direct as well as approximate methods to the uncorrected and the reference solutions. The settling velocity, $u_{s}$, based on Eq. \ref{eq:u_set_stokes} and the time scale of $d_p/u_s$ are employed for normalizing the results.}
    \label{fig:isotropic_normal}
\end{figure}

Figure \ref{fig:isotropic_normal} shows qualitatively the prediction of the different methods on particle velocity and trajectory of case N2 from Tab. \ref{tab:error_normal}. The settling velocity, $u_{s}$, based on Eq. \ref{eq:u_set_stokes} and the time scale of $d_p/u_s$ are employed for normalizing the results. As illustrated, the direct method captures the trajectory and velocity of the particle quite well in addition to the promising prediction of the approximate method compared to the uncorrected approach.

\subsection{Particle in oscillatory field}
As a final test case, the models are validated for unsteady motion of a single particle in an oscillatory flow field. A sinusoidal function is prescribed as a body force acting on the particle in an arbitrary direction of $i$, to resemble its unsteady motion in an oscillatory field. The following equation of motion is used for the studied cases of this part,

\begin{equation}
\frac{du_{i,p}}{dt} = \sin(\omega t)- \frac{f}{\tau_p}(u_{i,p}-u_{i,@p}^{ug}), 
\label{eq:oscillatory}
\end{equation}

\noindent with $f$ being calculated based on Eq. \ref{eq:schiller-Naumman} and $\omega$ being the frequency of the oscillation. The amplitude of oscillation is set to be unity for the sake of brevity. Similar to the previous sections, $u_{i,@p}^{ug}$ is the interpolated fluid velocity at the particle's location, that is incorrectly nonzero in the uncorrected two-way coupled simulations. For the reference, and consistent with other subsections, one-way coupled results are used wherein $u_{i,@p}^{ug}{=}0$. An additional non-dimensional parameter, Strouhal number, is also defined as, 

\begin{equation}
    Str = \frac{\tau^{e}_p}{\tau_w},
\end{equation}

\noindent which expresses the ratio of the particle time scale (Eq. \ref{eq:taup_eff}) and the time period of the oscillation, $\tau_w{=}1/\omega$. Table \ref{tab:oscillatory} lists the studied cases with the grid configurations employed in the previous subsections as well as various flow parameters. Defining a constant particle Reynolds number for this part might not be trivial due to the variation in the particle's velocity. Therefore, in order to set up the cases of this part, one can choose the maximum particle Reynolds number, $Re^{max}_p$, defined based on the maximum particle's velocity that occurs at the first crest point of its velocity profile. For the studied cases here, we perform two Strouhal number of $Str{=}0.1$ and $10$ and two maximum particle Reynolds number of $Re^{max}_p{=}0.097$ and $99.87$. Since finding an analytical expression for $Re^{max}_p$ might not be straightforward and setting up cases depends on this parameter as well, we provide the dimensional parameters that are needed for reproducing the reported cases here. The time step for the computations of this part, $\Delta t^{osc}$, follows the expression below which requires an additional constraint to the $\Delta t$ provided by Eq. \ref{eq:delta_t}, to accurately resolve the oscillation time scale, as well.  

\begin{equation}
    \Delta t^{osc} = {\rm min}\left(\Delta t, 0.03\tau_w\right)
\end{equation}

\begin{figure}[!htbp!]
    \centering
    \includegraphics[scale=0.55]{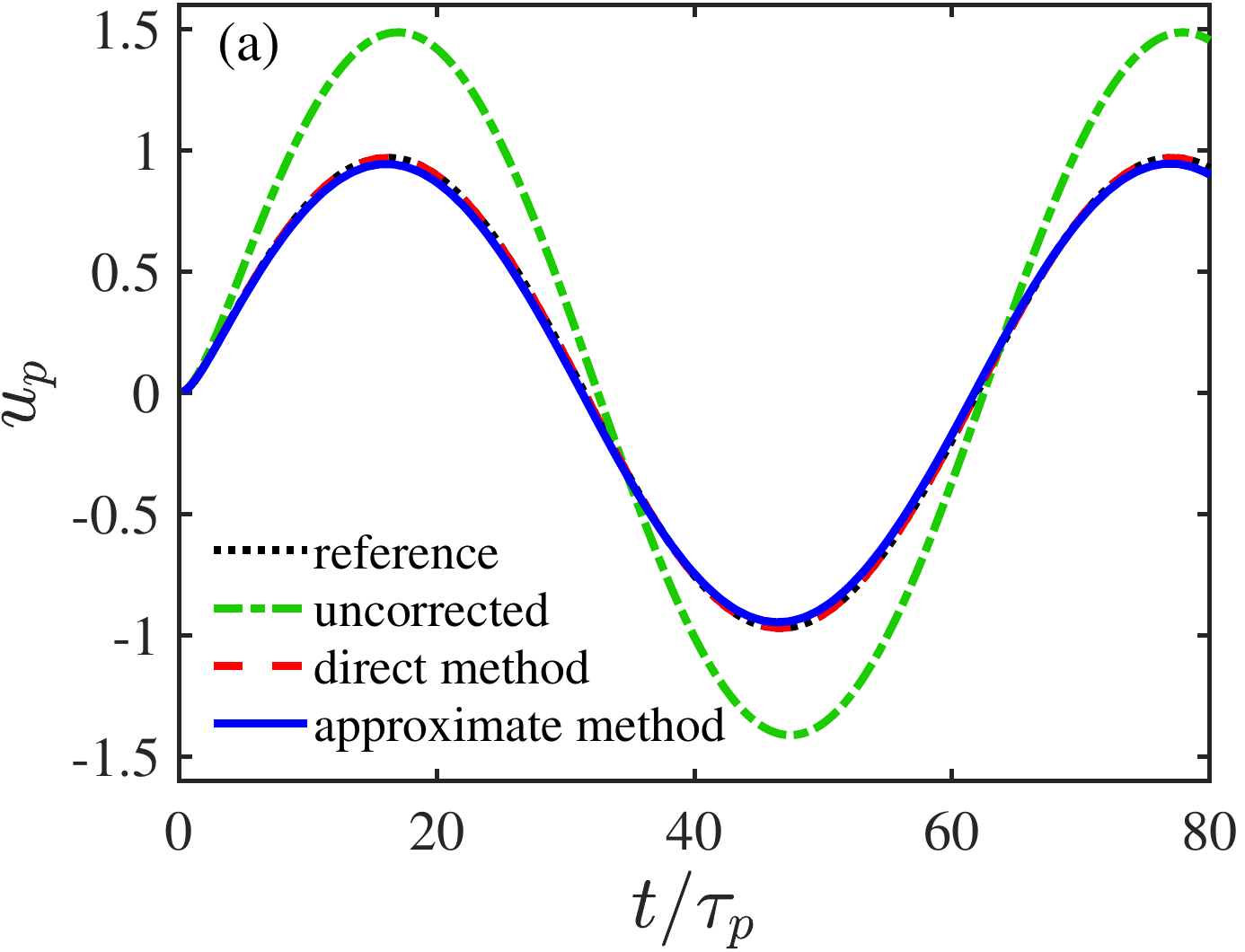}\hspace{0.03\textwidth}
    \includegraphics[scale=0.55]{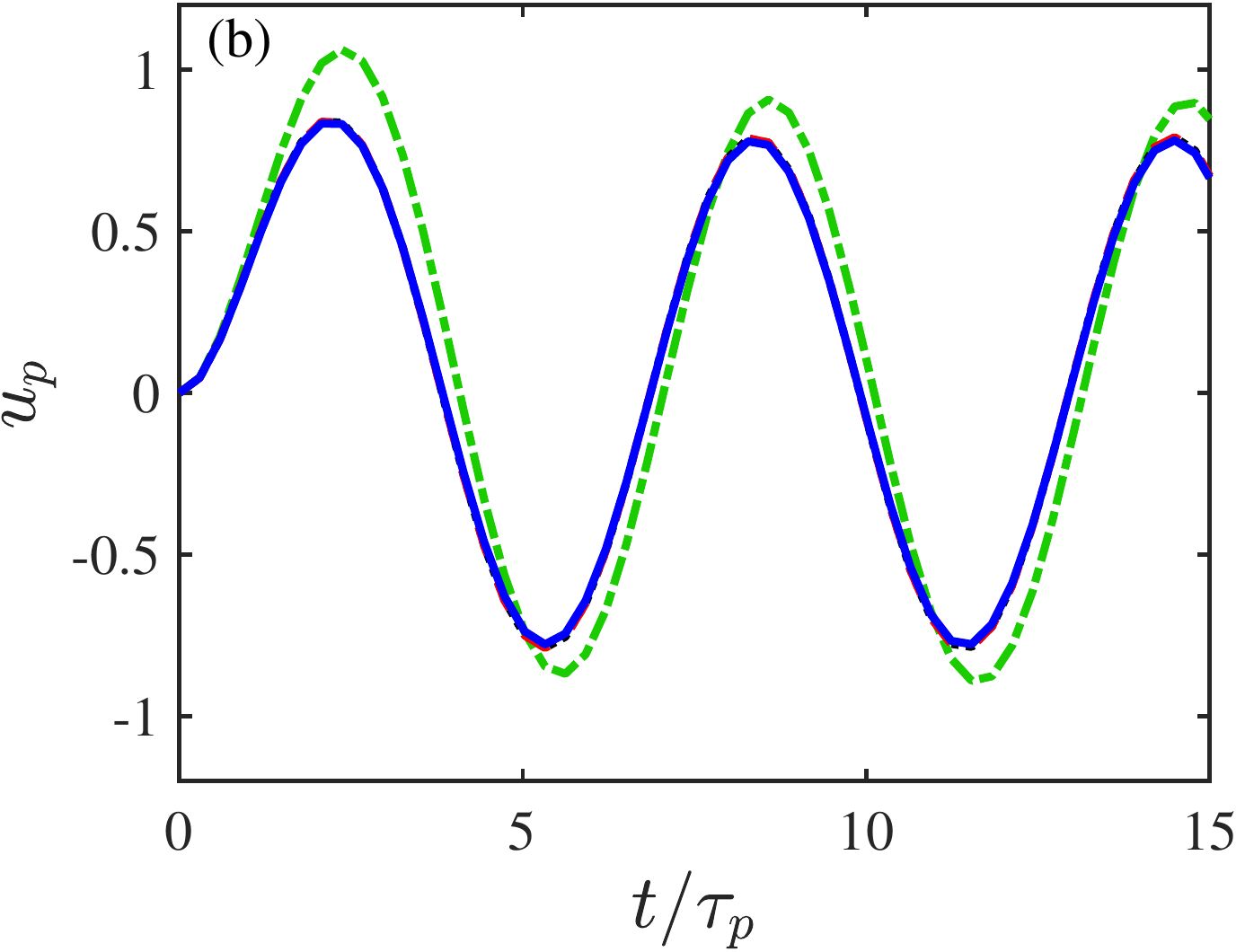}\vspace{0.03\textwidth}
    \includegraphics[scale=0.55]{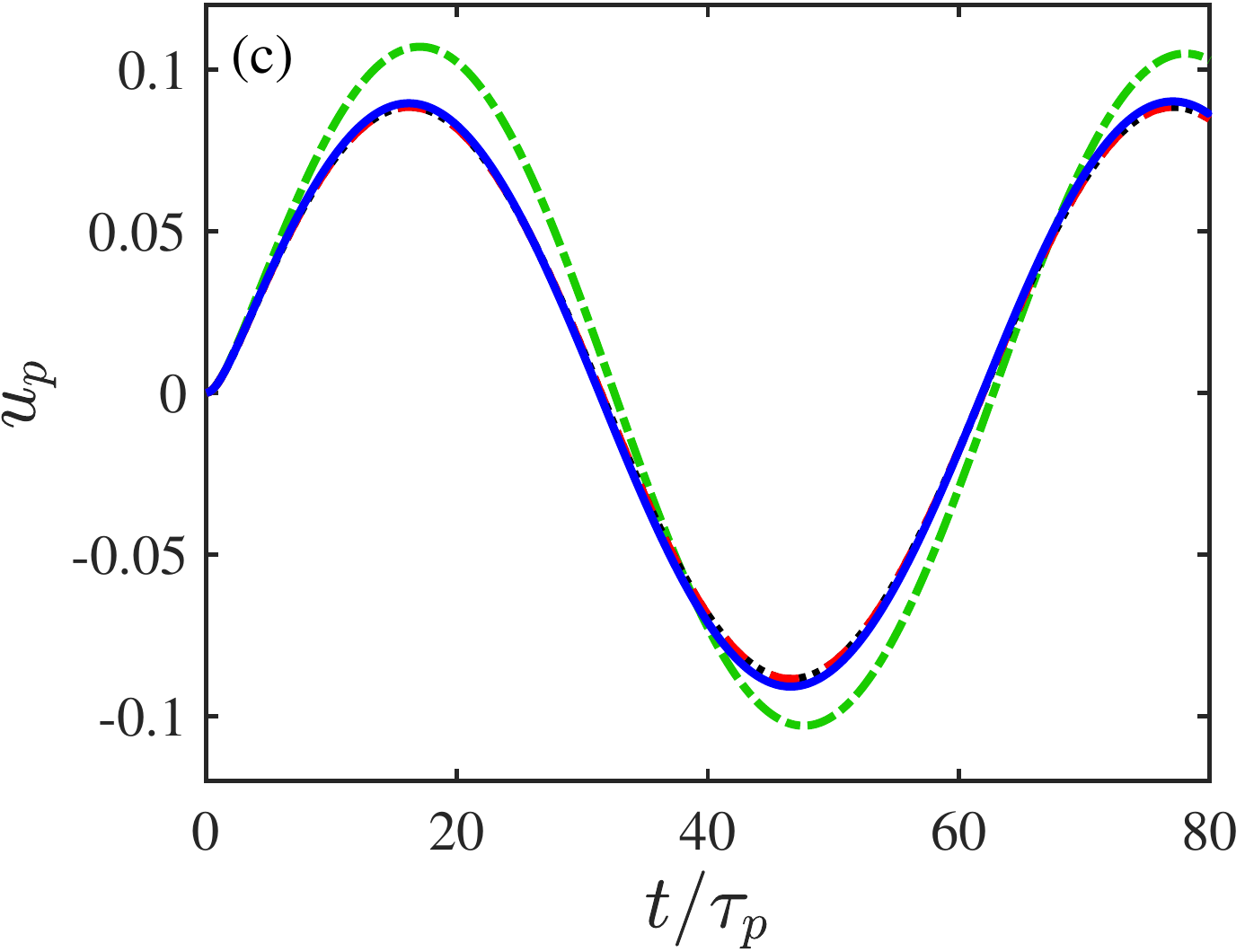}\hspace{0.03\textwidth}
    \includegraphics[scale=0.55]{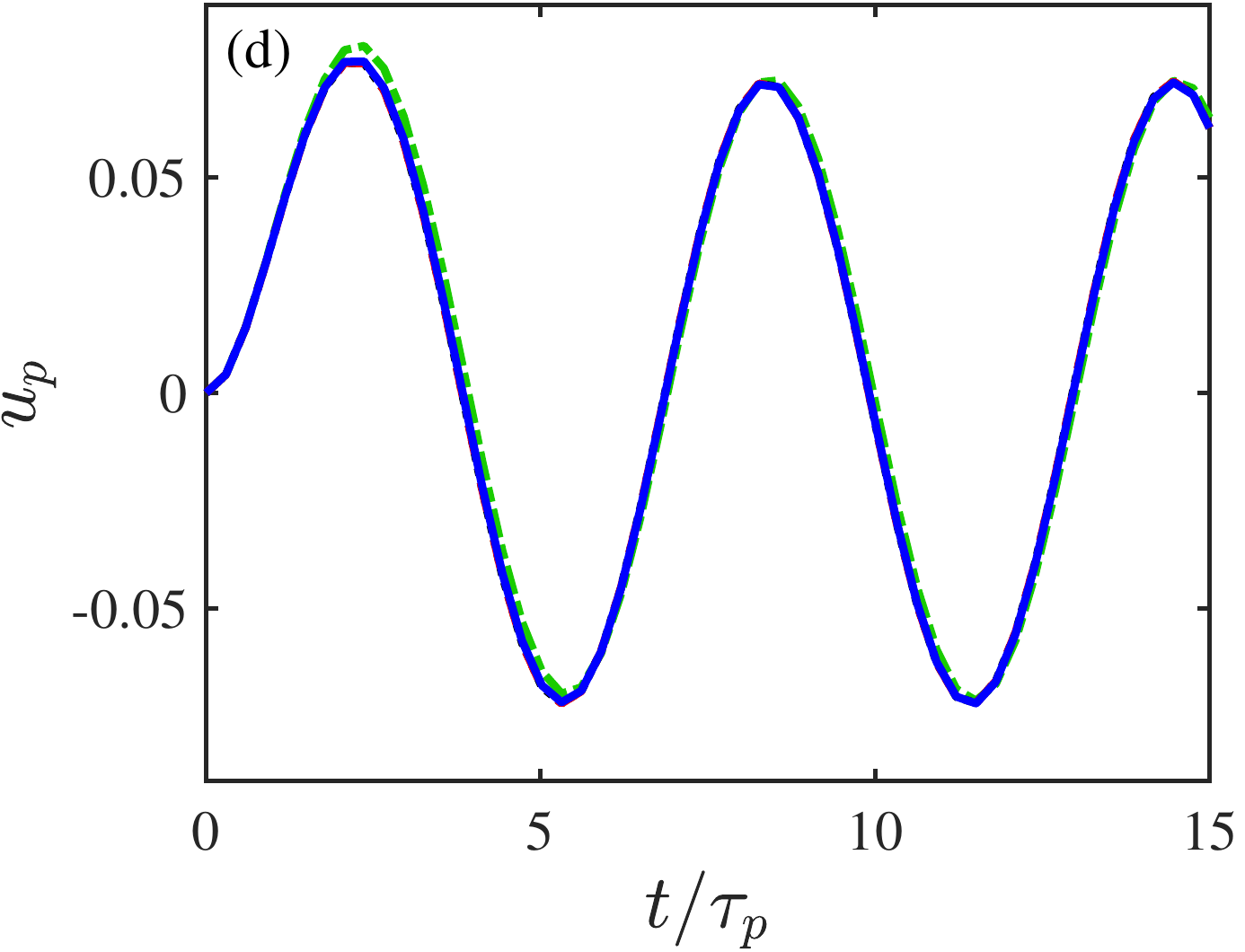}\vspace{0.03\textwidth}
    \includegraphics[scale=0.55]{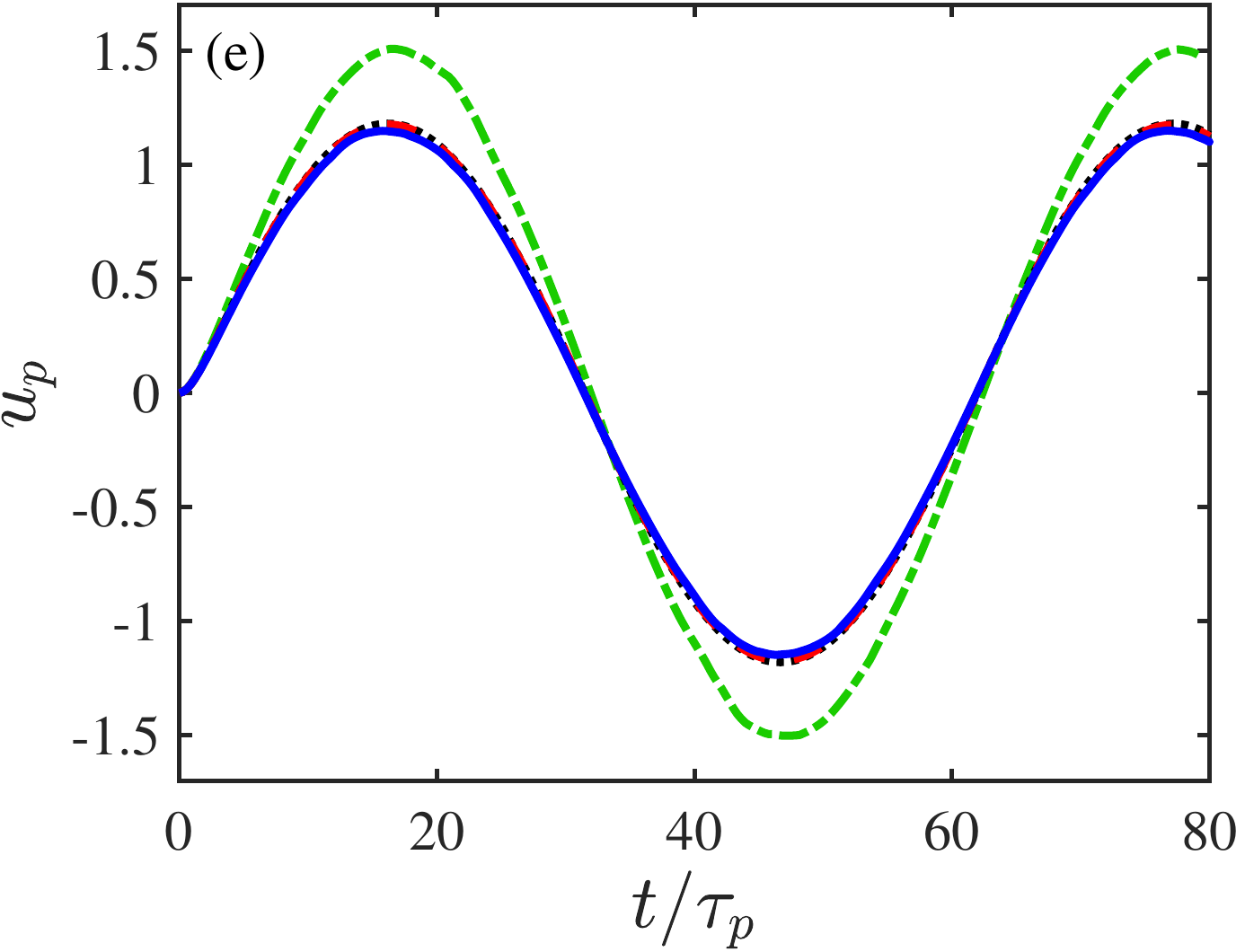}\hspace{0.03\textwidth}
    \includegraphics[scale=0.55]{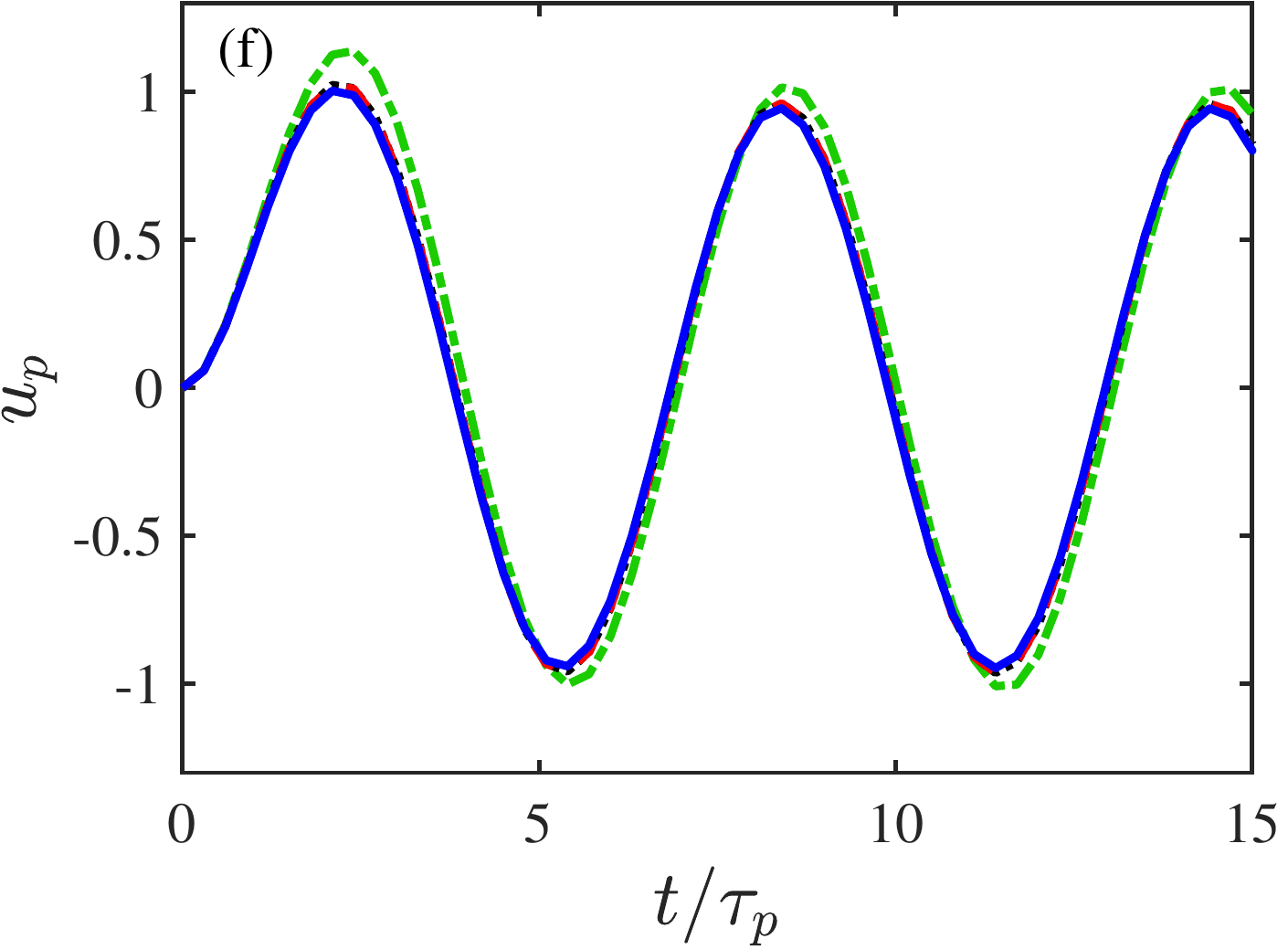}
    \caption{Time-dependent velocity [$m/s$] of a particle in an oscillatory motion is shown for the direct and approximate methods in comparison to the standard uncorrected scheme and the reference solution. Cases (a) to (f) pertain to O1, O2, O5, O6, O7 and O8, respectively, from Tab. \ref{tab:oscillatory}, that are all based on $Re^{max}_p{=}0.097$. Rows from top to bottom correspond to the cases pertaining to the isotropic rectilinear grid, anisotropic rectilinear grid and unstructured grid, respectively. Left and right columns pertain to cases with $Str{=}0.1$ and $Str{=}1.0$, respectively.}
    \label{fig:oscillatory_lowRe}
\end{figure}

\begin{figure}[!htbp!]
    \centering
    \includegraphics[scale=0.55]{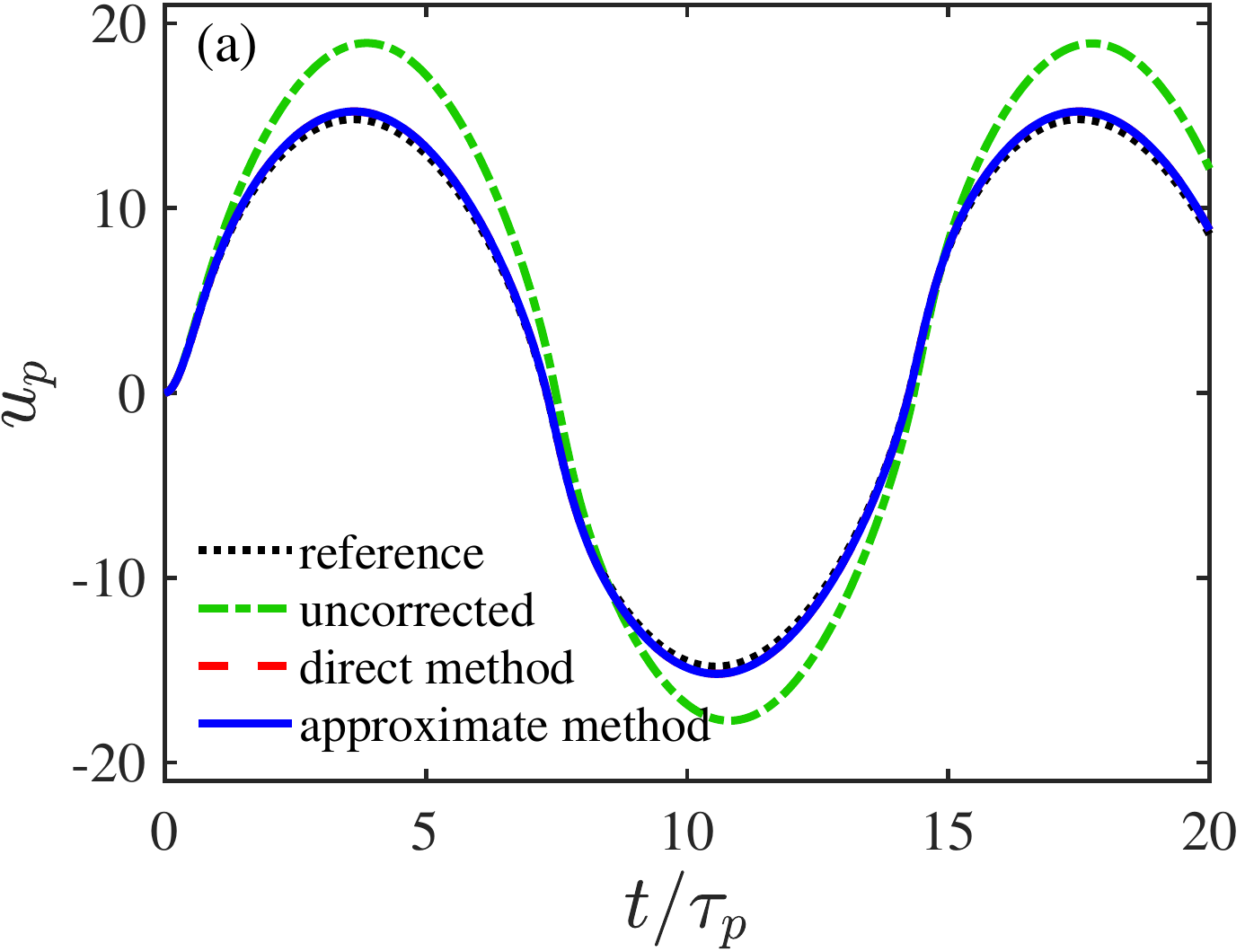}\hspace{0.035\textwidth}
    \includegraphics[scale=0.55]{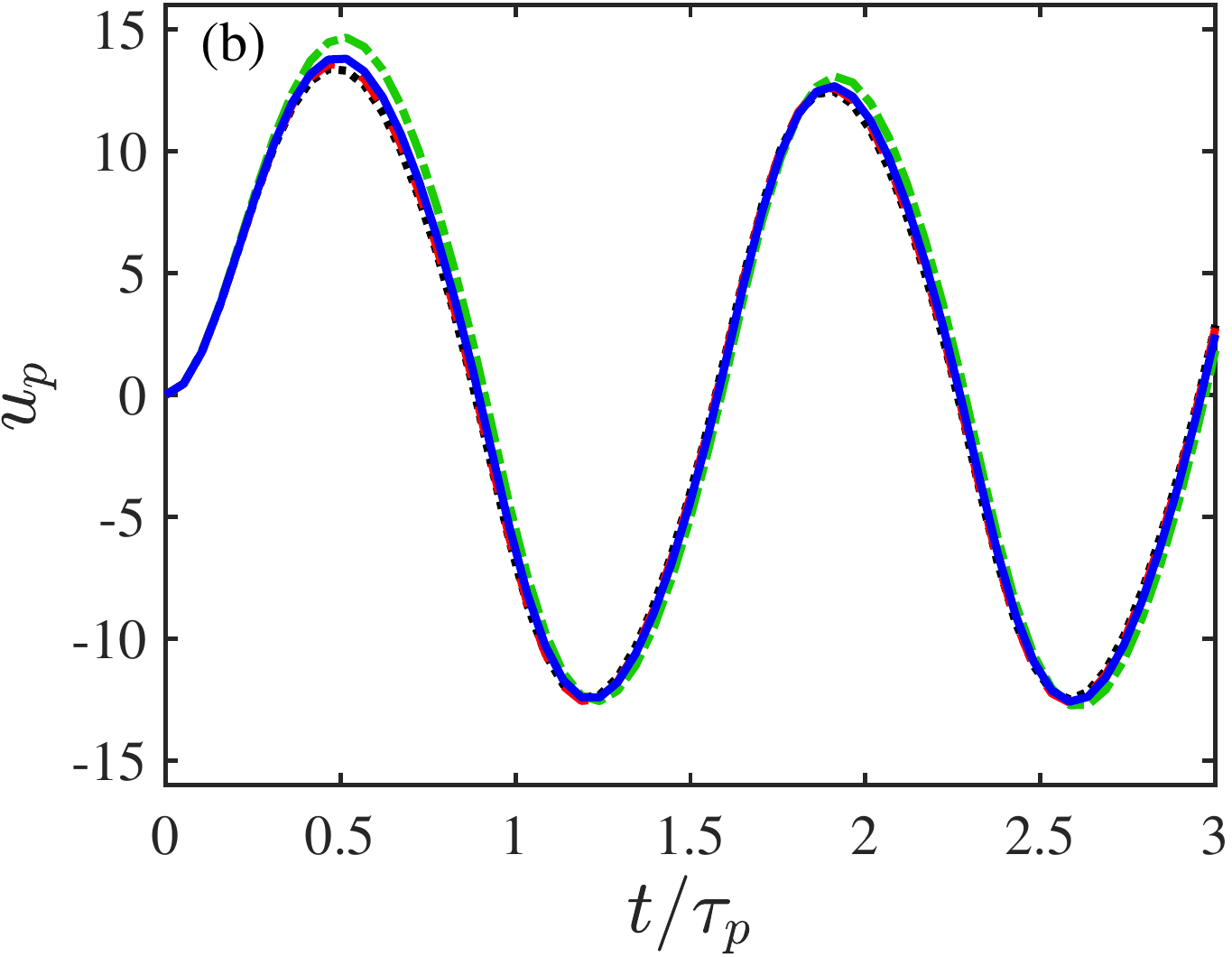}
    \caption{Time-dependent particle velocity [$m/s$] with $Re^{max}_p{=}99.87$ in the oscillatory field predicted by the present methods and uncorrected scheme in comparison to the reference. Results in (a) and (b) pertain to cases O3 and O4 from Tab. \ref{tab:oscillatory} with $Str{=}0.1$ and $Str{=}1.0$, respectively.}
    \label{fig:oscillatory_highRe}
\end{figure}

Figure \ref{fig:oscillatory_lowRe} shows excellent predictions of the present methods in capturing the time-dependent velocity of the particle in the unsteady field with $Re^{max}_p{=}0.097$ and different Strouhal numbers and using various grid configurations. As illustrated, the uncorrected scheme overshoots the crest and troughs of the particle velocity with significant deviation for small Strouhal number cases (left column of the figure), consistent with the observations of \cite{horwitz2016}. The performance of the present models for higher particle Reynolds numbers of $Re^{max}_p{=}99.87$, is shown in Fig. \ref{fig:oscillatory_highRe}, signifying the capability of the present correction methods even for unsteady motions, as well. 

\begin{table}
\begin{center}
\def~{\hphantom{0}}
\begin{adjustbox}{width=1.0\textwidth}
\begin{tabular} {lccccccccc}
\hline
cell shape & case & $Re^{max}_p$ & $Str$ & $\Lambda$ & $\omega(s^{-1})$ & $\mu (N.s/m^2)$ & $\rho_p(kg/m^3)$ & $d_p(m)$ & $\Delta t^{osc}(s)$ \\ 
\hline
&&&&&&&&& \\
\multirow{4}{*}{\includegraphics[scale=0.5]{./figures/cube.png}} & O1 & 0.097 & 0.1 & [1.0,1.0,1.0] & 0.1028 & 9.9700 & 180.0 & 1.0 & 0.0030 \\
& O2 & 0.097 & 1.0 & [1.0,1.0,1.0] & 0.8968 & 8.7000 & 180.0 & 1.0 & 0.0034\\
& O3 & 99.87 & 0.1 & [1.0,1.0,1.0] & 0.0067 & 0.1480 & 180.0 & 1.0 & 0.0400\\
& O4 & 99.82 & 1.0 & [1.0,1.0,1.0] & 0.0588 & 0.1292 & 180.0 & 1.0 & 0.0400 \\
&&&&&&&&& \\

\multirow{2}{*}{\includegraphics[scale=0.5]{./figures/rectangle.png}} & O5 & 0.097 & 0.1 & [0.3,6.0,0.6] & 1.1257 & 0.2730 & 5.0 & 0.3 & 0.0002\\
& O6 & 0.097 & 1.0 & [0.3,6.0,0.6] & 9.8136 & 0.2380 & 5.0 & 0.3 & 0.0003 \\
&&&&&&&&& \\

\multirow{2}{*}{\includegraphics[scale=0.5]{./figures/tetrahedral.png}} & O7 & 0.097 & 0.1 & 1.0 & 0.0844 & 17.99 & 180.0 & 1.482 & 0.0037 \\
& O8 & 0.097 & 1.0 & 1.0 & 0.7363 & 15.69 & 180.0 & 1.482 & 0.0042 \\
&&&&&&&&& \\

\hline
\end{tabular}
\end{adjustbox}
\caption{Listed are the cases performed for the unsteady motion of a particle in an oscillatory field. Two $Str$ and various grid configurations are performed for $Re^{max}_p{=}0.097$. For isotropic rectilinear grid, two additional cases with $Re^{max}_p{=}99.87$ are investigated, as well. $\rho_f{=}1.0 (kg/m^3)$ and $St{=}10.0$, are shared among all cases.}
\label{tab:oscillatory}
\end{center}
\end{table}

The test cases used in this study, underscore the applicability of the present models for a wide range of applications. Depending upon the accuracy necessary for simulation of particle-laden flows, either method can be chosen with the caveat that the direct method is more computationally expensive, but yields most accurate results. Concerning more sophisticated scenarios such as particles close to two different walls, curved walls, corners or rough walls, both methods are still applicable to recover the undisturbed fluid velocity as they solve momentum equations for the disturbance field using arbitrarily complex boundary conditions. However, solid conclusions on the capability of these methods in realistic and complex configurations are left for future works. Although the present work deals with incompressible flows, the concept of direct and approximate methods can be extended to compressible flows and variable density, reacting flows, as well as large-eddy simulation and Reynolds-averaged Navier-Stokes approaches.

\section{Conclusions}
\label{sec:conclusion}
A general, disturbance-corrected, point-particle (DCPP) formulation for two-way coupled computations of particle-laden flows is developed that recovers \color{black}the undisturbed flow field necessary for accurate computation of the fluid forces acting on the dispersed particles
The formulation is applicable to arbitrary shaped grids, both structured or unstructured, and in complex configurations involving no slip walls. To this end, governing equations for the disturbance created by the particle forces on the fluid flow are derived using the two-way coupled equations. 
Since the two-way coupled formulation and the disturbance flow equations are derived for any general boundary conditions, the developed approach is applicable to any complex flow with or without no-slip walls. The formulation can be implemented in any fluid flow solver and is not limited to specific types of grids.

Two models are developed to compute the disturbance field: (i) a direct method, and (ii) an approximate method. In the direct method, the non-linear disturbance momentum equations together with the continuity equation are solved using the same numerical formulation as the fluid flow solver for the two-way coupled field. This direct method provides the disturbance velocity and pressure fields, is free of any empirical or calibrated expressions, and makes it attractive for a wide range of computations including complex geometries, arbitrary shaped unstructured grids, as well as particle-laden flows with any arbitrary particle size and particle Reynolds number. However, the cost associated with this approach is nearly doubled, as the disturbance momentum together with the continuity constraint require additional Poisson solution for the disturbance pressure. Nevertheless, the accuracy gained by such computation warrants its use, and the cost is still considerably lower than particle-resolved, direct numerical solutions wherein the grid resolutions used are much finer than the particle size.

In order to alleviate the computational cost associated with the direct method, a reduced order, approximate method was introduced, wherein a simplified momentum equation for the disturbance field is solved. This approximate model is based on low Reynolds number assumption and neglects the non-linear, advective terms. In addition, in the steady, Stokes flow limit, the Stokes solution over a spherical particle motivates an approximation for the pressure gradient term. 
The pressure and viscous terms are modeled by a modified viscous term with an effective increased viscosity determined to match pressure and viscous forces in the Stokes flow limit. Since, the pressure field is not directly computed, the continuity constraint is only indirectly imposed, and the expensive step of solving a Poisson equation for the disturbance field is not needed.
Although the approximate model was constructed based on the assumption of small $Re_p$, where inertial effects are negligible, the test cases for high particle Reynolds numbers, up to $Re_p{=}100$, remarkably show good predictive capability of this approach. In addition, the accuracy of this approximate method can be further improved by making the effective viscosity a function of the particle Reynolds number, however, for majority of the cases studied, this was not necessary.
The approximate method is shown to be as accurate as the direct method and has considerable reduced computational cost that is on the same order of the existing correction schemes.

Both models were tested for various scenarios using isotropic and anisotropic rectilinear grids, tetrahedral unstructured grid, different particle sizes, a wide range of particle Reynolds numbers ($0{\leq}Re_p{\leq}100$), different particle-to-grid size ratios, $\Lambda$, and in the presence and absence of no-slip walls. Both methods showed excellent predictions of particle settling velocity in an unbounded regime with small errors in settling and drift velocities. Errors in the uncorrected scheme were significant for large $\Lambda$ and small $Re_p$ with the fact that the need for correction diminishes when particle settling Reynolds number increases~\citep{balachandar2019,pakseresht2020}.

Prediction of particle settling near a no-slip wall was also evaluated for a single particle in parallel and normal motion to the wall.
It was shown that both the direct and approximate methods were capable of recovering the undisturbed field and reduced the errors to small values for particle settling parallel to a no-slip wall. 
For particle motion normal to a wall, the direct method showed excellent prediction in recovering the undisturbed field and produced correct trajectory and velocity of the particle. The approximate method also produced small errors for nearly isotropic grids; even for particles larger than the grid resolution. However, for highly skewed anisotropic grids, its overprediction in the undisturbed field yielded errors on the same order of magnitude as the uncorrected scheme. An interpolation stencil that scales with the particle size, may alleviate this issue. In addition, for particle motion normal to a no-slip wall, the pressure distribution on the particle surface is asymmetric, and thus a simple approach to model the pressure gradient used in the approximate method potentially needs to be modified. Nevertheless, the approximate method is capable of capturing motion of a particle near a wall accurately, especially for nearly isotropic and arbitrary shaped grids.

Finally, to test the models for unsteady motion of particles, an inevitable criterion in complex particle-turbulence interaction, particle in oscillatory motion was investigated varying the Strouhal number ($0.1$ and $1.0$), the ratio of the oscillation time scale and particle relaxation time, and two particle Reynolds numbers approximately $0.1$ and $100.0$. Excellent predictions were achieved using both methods revealing their predictive capability even in unsteady motion. 

The present DCPP approach can be easily implemented in Euler-Lagrange and Euler-Euler packages as it leverages the identical algorithm, boundary conditions, and type of the computational gird that are employed for solving the standard two-way coupled PP approaches. In its current form, the DCPP approach is directly applicable to systems with dilute volume loading, wherein the particle-particle hydrodynamic interactions are negligible. The approach can also be applied to dense regimes, by explicitly modeling the hydrodynamic interaction and neighboring particle effects.

\section{Acknowledgements}
Financial support was provided under the NASA Contract Number NNX16AB07A monitored by program manager Dr. Jeff Moder, NASA Glenn Research Center as well as the National Science Foundation (NSF) under Grant Number 1851389. In addition, the authors acknowledge the Texas Advanced Computing Center (TACC) at the University of Texas at Austin for providing HPC resources that have contributed to the results reported here. The authors acknowledge Mr. Shashank Karra for generating the unstructured grids for various configurations used in the present work. 


\section*{Appendix A. Numerical formulation}
The numerical approach used is based on fractional time-stepping on colocated, arbitrary shaped, unstructured, grid elements for constant density, incompressible flows.  A semi-implicit scheme is used for the momentum equation solution, however, the inter-phase momentum exchange terms are treated explicitly. The collocated grid arrangement is used for its easy application to structured as well as arbitrary unstructured grids.

\begin{figure}[!htpb!]
    \centering
    \includegraphics[scale=0.5]{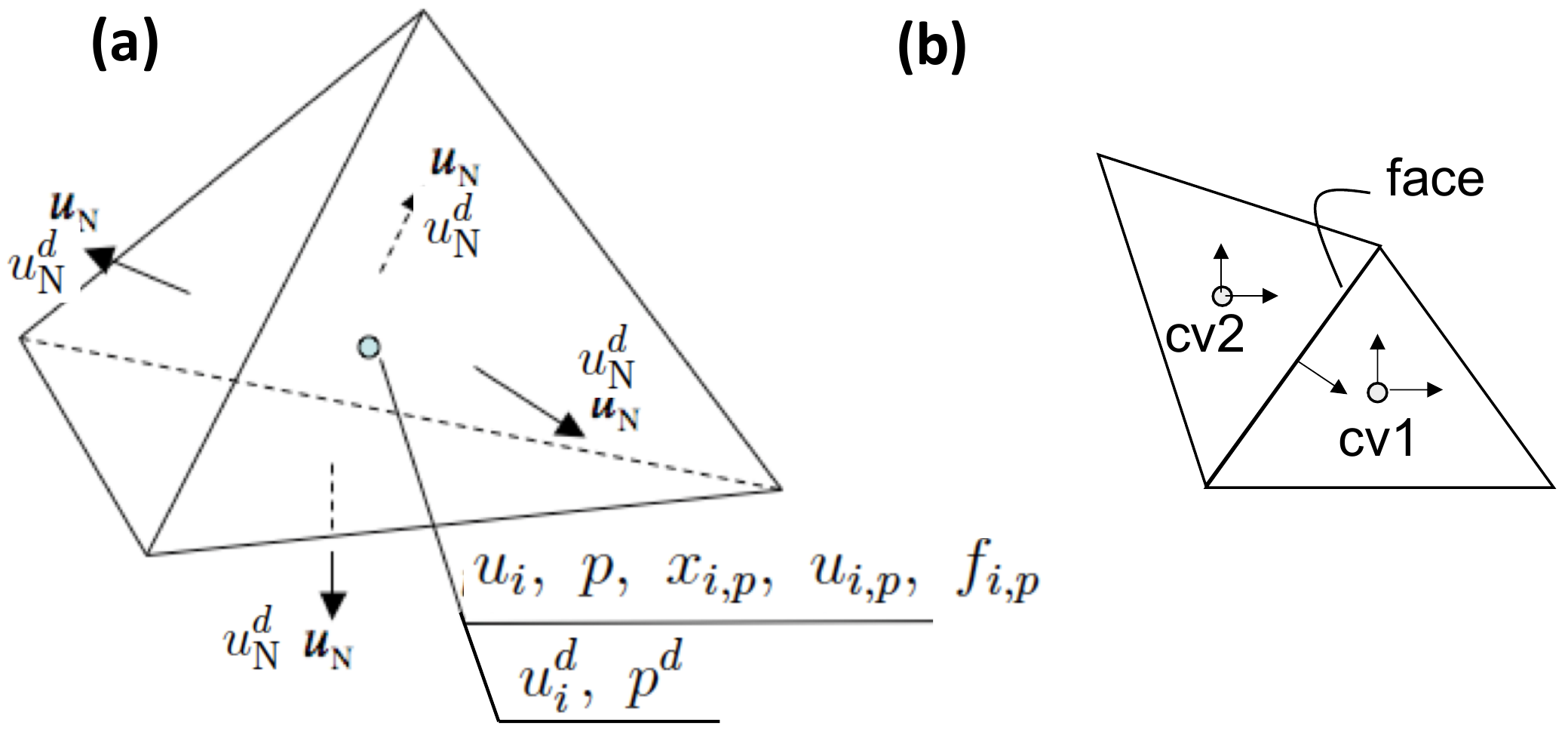}
    \caption{Stencil used for the colocated grid, fractional time-stepping based algorithm for the two-way coupled flow field ($u_i$, $p$) and the disturbance field ($u_i^d$, $p^d$).}
    \label{fig:stencil}
\end{figure}

Figure~\ref{fig:stencil} shows the schematic of variable storage for fluid and particle phases.
All variables are stored at the control volume (cv) center with the exception of the face-normal velocity $u_{\rm N}$ (and $u_{\rm N}^d$), located at the face centers. The face-normal velocity is used to enforce continuity equation. Subscript `p' is used to denote the disperse phase. Using these variable locations, integrating the governing equations over the control volume and applying Gauss' divergence theorem to convert volume integrals to surface integrals wherever possible, the discrete governing equations are derived. Accordingly, the continuity equation is
\begin{equation}
 \frac{1}{{\mathcal V}_{\rm cv}}\sum_{\rm faces~of~cv} u_{\rm N}^{n+1} A_{\rm face} = 0,
\end{equation}
where $\Delta t$ is the flow solver time-step, ${\mathcal V}_{\rm cv}$ is the volume of the cv, $A_{\rm face}$ is the area of the face of a cv and $u_{\rm N}$ is the face-normal velocity. For the present colocated grid finite volume scheme the face-normal velocity $u_{\rm N}$ is obtained through a projection scheme rather than interpolation of the control volume based velocity to the faces. The discrete momentum equation for the $i$ component of velocity can be written as
\begin{eqnarray*}
\frac{g_{i,\rm cv}^{n+1} - g_{i,\rm cv}^n}{\Delta t}  +
\frac{1}{{\mathcal V}_{\rm cv}}\sum_{\rm faces~of~cv}g_{i,\rm face}^{n+1/2}u_{\rm N}^{n+1/2}A_{\rm face} &=&
        -\frac{\partial}{\partial x_i} p_{\rm cv}^{n+1} +
\end{eqnarray*}
\begin{equation}
~~~~~~~~~~~~
      \frac{1}{{\mathcal V}_{\rm cv}}\sum_{\rm faces~of~cv}\left(\tau_{ij}\right)_{\rm face}^{n+1/2}{\rm N}_{j,\rm face}A_{\rm face}
        + {f}_{i,\rm cv}^{n+1},
\end{equation}
where $g_{i}=\rho_f u_{i}$ represents the momentum per unit volume in the $i$ direction, $\rho_f$ is the constant fluid density, $(\tau_{ij})_{\rm face}$ is the viscous stress at the faces of control volume, and ${\rm N}_{j,\rm face}$ represents
the components of the outward face-normal. Similarly, the velocity field ($u_{i,\rm face}$), and the momentum $g_{i,\rm face} = \rho_f u_{i,\rm face}$ at the faces are obtained using arithmetic averages of the corresponding fields at the two control volumes associated with the face. The values at time level $t^{n+1/2}$ are obtained by simple time-averaging. The interface coupling force is represented by ${f}_{i,\rm cv}$. The pressure field $p_{\rm cv}^{n+1}$ is unknown and is obtained using the best available guess at the current iteration. This gets updated during the solution of the pressure Poisson equation. 
The above discretization is implicit and thus the time-steps are not limited by viscous stability limits. The use of symmetric centered differences makes the algorithm second order on uniform Cartesian grids.
The main steps of the solver are described below.
\begin{itemize}
\item{{\bf Step 1:}
Set the flow velocity at $t^{n+1}$ using a second-order Adams-Bashforth predictor. Advance the particle positions and velocities using the undisturbed fluid velocity obtained from the solution of the two-way coupled fluid flow equations ($u_i$ and $p$) and the disturbance velocity and pressure fields ($u_i^d$ and $p^d$) from the previous time step. 
}

\item{{\bf Step 2:}
Advance the two-way coupled fluid momentum equations using the fractional step algorithm, with the interphase force, $f_i$, treated explicitly.
\begin{eqnarray}
\label{eq:mom2_num}
\rho_f \frac{u_i^{*} - u_i^n}{ \Delta t} + 
\frac{\rho_f}{ 2{\mathcal V}_{\rm cv}}  \sum_{\rm faces~of~cv}\left[u_{i,\rm face}^{*} + u_{i,\rm face}^n\right]u_{\rm N}^{n+1/2}A_{\rm face} = \nonumber \\
-\frac{\delta p}{\delta x_i}^{n}+\frac{\mu}{ 2{\mathcal V}_{\rm cv}}\sum_{\rm faces~of~cv}\left(\frac{\partial u_{i,\rm face}^{*}}{ \partial x_j}
+ \frac{\partial u_{j,\rm face}^n}{\partial x_i} \right)A_{\rm face} +  f_i^{n+1}, 
\end{eqnarray}
where $\rm N$ is the face-normal component, and $A_{\rm face}$ is the face area, $\mu$ is the fluid viscosity, and $\rho_f$ the density. The pressure gradient at the CV centers in the above equation is at the old time-level and is obtained as described below in Step 6. The reaction force $f_{i}^{n+1}$ is obtained through Eulerian-Lagrangian interpolation (equation~\ref{eq:delta}). In the above step, the viscous terms are treated implicitly, the three equations for the velocity components at the CV centers are solved using iterative scheme such as Gauss-Seidel or algebraic multigrid solvers.
}
\item{{\bf Step 3:} Remove the old pressure gradient to obtain the velocity field, $\widehat u_{i}$:
\begin{equation}
\frac{{\widehat u_i} -  u_i^{*}}{\Delta t}  = + \frac{1}{\rho_f} \frac{\delta p}{\delta x_i}^n
\end{equation}
}
\item{{\bf Step 4:}
Interpolate the velocity fields to the faces of the control volumes and consider the corrector step:
\begin{eqnarray}
\rho_f \frac{u_{\rm N}^{n+1}-{\widehat u}_{\rm N}}{\Delta t} &=& 
	-\frac{\delta p}{\delta x_{\rm N}}^{n+1},
\end{eqnarray}
where ${\widehat u}_{\rm N} = {\widehat u}_{i,\rm face} {\rm N}_{i,\rm face}$
is the approximation for face-normal
velocity and ${\rm N}_{i,\rm face}$ are the components of the face-normal. The face-based velocity is simply obtained as the average of the two control volumes that share the common face, ${\widehat u}_{i,\rm face} = ( {\widehat u}_{i,{\rm cv1}} + {\widehat u}_{i,{\rm cv2}})/2$ as shown in figure~\ref{fig:stencil}b. To face-based pressure gradient also makes use of the two adjacent cvs:
\begin{equation}
\frac{\delta p^{n+1}}{\delta x_{\rm N}} = \frac{ p_{\rm nbr}^{n+1} - p_{\rm cv}^{n+1}} {|{\mathbf S}_{\rm cv\rightarrow nbr}|},
\end{equation}
where the subscripts `$\rm cv$' and `$\rm nbr$' stand for the the control volume for which the velocity field is being solved and the neighboring control volumes sharing a common face, respectively and $|{\mathbf S}_{\rm cv\rightarrow nbr}|$ represents the magnitude of the position vector connecting the two control volumes. 
}
\item{{\bf Step 5:}
 The pressure field and the pressure gradients at $t^{n+1}$ are unknown in the above step. A pressure Poisson equation is derived by taking a discrete divergence of the above equations and solving for the pressure field at each control volume:
\begin{equation}
\label{eq:Poisson}
\sum_{\rm face~of~cv} \frac{\delta p^{n+1}}{\delta x_N} = 
	\frac{\rho_f}{\Delta t}\sum_{\rm faces~of~cv}{\widehat u}_{i,\rm face} A_{\rm face} 
\end{equation}
}
\item{{\bf Step 6:}
Reconstruct the pressure gradient at the cv-centers. The
face-normal pressure gradient $\delta p/\delta x_N$ 
and the gradient in pressure at the cv-centers are related by the area-weighted
least-squares interpolation~\citep{mahesh2004numerical}:
\begin{equation}
\epsilon_{cv} = \sum_{\rm faces~of~cv} \left(P^{\prime}_{i,\rm cv}{\rm N}_{i,\rm face} - P^{\prime}_{\rm face}\right)^2A_{\rm face},
\end{equation}
where $P^{\prime}_{i,\rm cv} = \frac{\delta p}{\delta x_i}$ and $P^{\prime}_{\rm face} = \frac{\delta p}{\delta x_{\rm N}}$. 
}
\item{{\bf Step 7:}
Compute new face-based velocities, and update the cv-velocities:
\begin{eqnarray}
u_{\rm N}^{n+1} &=& {\widehat u}_{\rm N} - \frac{\Delta t }{ \rho_f} \frac{\delta p^{n+1}}{\delta x_{\rm N}} \\
u_{i,\rm cv}^{n+1} &=& {\widehat u}_{i,\rm cv} - \frac{\Delta t }{\rho_f} \frac{\delta p^{n+1}}{ \delta x_{i, \rm cv}}
\end{eqnarray}
}
\item{{\bf Step 8:} Repeat steps 2--7 for the disturbance field, $u_i^d$ and $p^d$, solving the momentum and continuity equations expressed by Eqs.~\ref{eq:moment_dist} and \ref{eq:cont_dist} for the direct method. For the approximate method, repeat step 2 solving the approximate disturbance equation provided by Eq.~\ref{eq:moment_appx}.
}

\item{{\bf Step 9:} Using the disturbance field, compute the undisturbed flow velocity (direct and approximate method) and pressure (direct method) and proceed to Step 1.
}

\end{itemize}

\bibliographystyle{elsarticle-harv}\biboptions{authoryear}
\bibliography{paksereshtapte2020}

\end{document}